\documentclass[revtex4-1,twocolumn,superscriptaddress]{revtex4-1}

\usepackage[utf8]{inputenc}
\usepackage{graphicx}  
\usepackage{amsmath,mathtools}
\usepackage{amssymb}
\usepackage{enumerate}
\usepackage{dsfont}
\usepackage[toc,page]{appendix}
\pdfoutput=1

\newcommand{\Eq}[1]{Eqn.~\ref{#1}}
\newcommand{\Eqs}[1]{Equations~\ref{#1}}

\newcommand{\Fig}[1]{Fig.~\ref{#1}}

\newcommand{\be}{\begin{enumerate}}
\newcommand{\ee}{\end{enumerate}}
\newcommand{\bi}{\begin{itemize}}
\newcommand{\ei}{\end{itemize}}

\newcommand{\II}{\mathbb{I}}

\newcommand{\dif}{\mathrm{d}}

\newcommand{\cross}{\times}

\newcommand{\diagentry}[1]{\mathmakebox[1.8em]{#1}}
\newcommand{\xddots}{%
  \raise 4pt \hbox {.}
  \mkern 6mu
  \raise 1pt \hbox {.}
  \mkern 6mu
  \raise -2pt \hbox {.}
}

\usepackage{hyperref}
\makeatletter
\newcommand{\customlabel}[2]{%
   \protected@write \@auxout {}{\string \newlabel {#1}{{#2}{\thepage}{#2}{#1}{}} }%
   \hypertarget{#1}{}
}
\makeatother

\makeatletter
\newsavebox\myboxA
\newsavebox\myboxB
\newlength\mylenA
\newcommand*\xoverline[2][0.75]{%
    \sbox{\myboxA}{$\m@th#2$}%
    \setbox\myboxB\null
    \ht\myboxB=\ht\myboxA%
    \dp\myboxB=\dp\myboxA%
    \wd\myboxB=#1\wd\myboxA
    \sbox\myboxB{$\m@th\overline{\copy\myboxB}$}
    \setlength\mylenA{\the\wd\myboxA}
    \addtolength\mylenA{-\the\wd\myboxB}%
    \ifdim\wd\myboxB<\wd\myboxA%
       \rlap{\hskip 0.5\mylenA\usebox\myboxB}{\usebox\myboxA}%
    \else
        \hskip -0.5\mylenA\rlap{\usebox\myboxA}{\hskip 0.5\mylenA\usebox\myboxB}%
    \fi}
\makeatother

\newcommand{\eq}{\begin{equation}}
\newcommand{\qe}{\end{equation}}

\usepackage{color}

\begin{document}

\title{Origin and localization of topological band gaps in gyroscopic metamaterials
}

\author{Noah P. Mitchell}
\email{npmitchell@kitp.ucsb.edu}
\thanks{Corresponding author}
\affiliation{Kavli Institute for Theoretical Physics, University of California Santa Barbara}
\affiliation{James Franck Institute and Department of Physics, University of Chicago, Chicago, IL 60637, USA}
\author{Ari M. Turner}
\email{turner.ari@ph.technion.ac.il}
\thanks{Corresponding author}
\affiliation{Department of Physics, Israel Institute of Technology}
\author{William T. M. Irvine}
\email{wtmirvine@uchicago.edu}
\thanks{Corresponding author}
\affiliation{James Franck Institute and Department of Physics, University of Chicago, Chicago, IL 60637, USA}
\affiliation{Enrico Fermi Institute, The University of Chicago, Chicago, IL 60637, USA}

\begin{abstract}
Networks of interacting gyroscopes have proven to be versatile structures for understanding and harnessing finite-frequency topological excitations.
Spinning components give rise to band gaps and topologically protected wave transport along the system's boundaries, whether the gyroscopes are arranged in a lattice or in an amorphous configuration. 
Here, we examine the irrelevance of periodic order for generating topological gaps.
Starting from the symplectic dynamics of our model metamaterial, we present a general method for predicting whether a gap exists and for approximating the Chern number using only local features of a network, bypassing the costly diagonalization of the system's dynamical matrix.
We then study how strong disorder interacts with band topology in gyroscopic metamaterials and find that amorphous gyroscopic Chern insulators exhibit similar critical behavior to periodic lattices.
Our experiments and simulations additionally reveal a topological Anderson insulation transition, wherein disorder drives a trivial phase into a topological one.


\end{abstract}


\maketitle


\section{Introduction}

The past decade has witnessed a surge of interest in exploiting concepts from topology as a powerful tool to engineer material behavior~\cite{kosterlitz_ordering_1973,haldane_model_1988,sanchez_spontaneous_2012,coulais_static_2017}.
In particular, an important realization was that collections of a material's normal modes are not merely independent; instead, phase relationships between modes -- quantified, for example, by Berry curvature -- can have physical implications.  
This realization provides a natural framework for linking topological ideas to material systems and distinguishing topologically distinct phases of matter~\cite{laughlin_quantized_1981,thouless_quantized_1982,prodan_topological_2009}. 


These connections have revealed unique phenomena governed by the topology of the band structure of the material in question.
Crucially, the resulting exotic effects, such as the appearance of chiral edge modes, are protected from disorder by their topological origin, meaning that chiral edge modes are insensitive to geometric features of the boundary or inhomogeneity in the material's bulk~\cite{hasan_colloquium:_2010,ryu_topological_2010}.
The design of mechanical metamaterials -- engineered structures assembled from composite components -- has extended topological protection to elastic structures, enabling us to control and direct phononic excitations robustly without requiring extreme precision in the metamaterial's construction~\cite{prodan_topological_2009,kane_topological_2013,susstrunk_observation_2015}.

At the same time, metamaterials offer an ideal arena for fundamental insights into topological phases. 
In metamaterials, we observe topological excitations directly at a macroscopic scale, rather than through indirect transport measurements.
Tabletop metamaterials furthermore support rapid prototyping and access to the material's individual components.
This class of platforms has allowed the direct identification of topological behavior in crystalline systems as diverse as coupled pendula~\cite{susstrunk_observation_2015}, photonic waveguides~\cite{rechtsman_photonic_2013}, isostatic frames~\cite{paulose_selective_2015}, chains of gears~\cite{meeussen_geared_2016}, and lattices of acoustic resonators ~\cite{souslov_topological_2017}. 
One other emerging platform of metamaterial systems is composed of interacting gyroscopes suspended from a plate~\cite{nash_topological_2015,mitchell_realization_2018,mitchell_tunable_2018,wang_topological_2015,brun_vortex-type_2012,carta_dispersion_2014,carta_flexural_2019,susstrunk_classification_2016}. 
Because networks of gyroscopes break time reversal symmetry in a geometry-dependent manner, these gyroscopic tabletop metamaterials have provided a simple platform to observe topological phase transitions~\cite{mitchell_realization_2018}
and switch between multiple topological phases with multiple gaps~\cite{mitchell_tunable_2018}. 

This ease of access in turn allows for discovery.
For example, constructing amorphous materials using the gyroscopic platform provided a route to discovering that topology arises not just in crystalline materials, but even in amorphous networks~\cite{mitchell_amorphous_2018}.
This last example suggests that topological phases should arise in glassy systems composed of randomly arranged inclusions, dopants, or vacancies -- a prediction that has since been identified in models of quantum matter~\cite{agarwala_topological_2017,sahlberg_topological_2020,marsal_topological_2020,poyhonen_amorphous_2018,huang_theory_2018,agarwala_higher-order_2020}.
Some instances of these topological amorphous phases have been suggested to be distinct phases of matter~\cite{bourne_non-commutative_2018,sahlberg_topological_2020} and have been placed into a topological classification scheme for mechanical systems~\cite{barlas_topological_2018}.

\begin{figure}
    \centering
    \includegraphics[width=\linewidth]{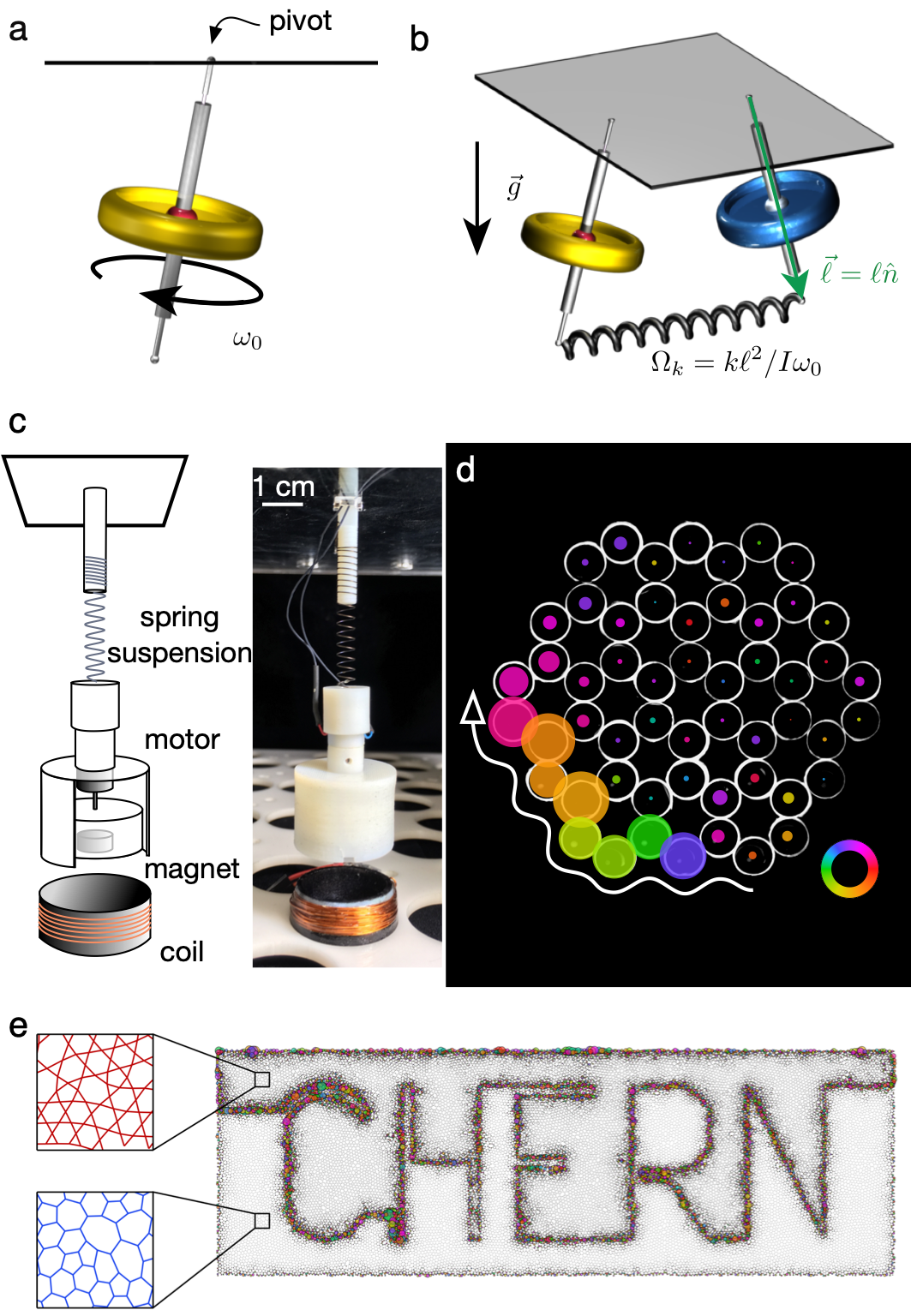}
    \caption{
    \textbf{Gyroscopic metamaterials provide a platform for engineering exotic behavior.}
    \textit{(a)} Each gyroscope is suspended from a plate and spins rapidly with a speed $\omega_0$. 
    \textit{(b)} 
    Coupled gyroscopes experience an interaction frequency, $\Omega_k$, and precess under the influence of restoring forces such as gravity.
    \textit{(c)} In an experimental realization, externally-powered motors spin small masses with magnets embedded at their cores. 
    All gyroscopes spin in the same direction.
    A weak spring suspends the gyroscope from a plate, enabling low-loss precession. 
    A coil beneath each gyroscope locally modulates the effective gravitational torque on each gyroscope, allowing us to introduce disorder or staggered potentials.
    \textit{(d)} Chiral waves skim along the boundary when the system is shaken at a frequency in the resulting bandgap. 
    Each colored circle represents the measured displacement of a gyroscope in our experimental realization, with the radius proportional to the displacement amplitude and the color denoting the phase (inset shows colorwheel).
    \textit{(e)} Protected chiral waves persist even if the spatial structure of the underlying network is amorphous. 
    Here, a simulated amorphous network of gyroscopes consists of a kagome-like structure above and a Voronoi-like structure below a boundary shaped into the letters `CHERN'. 
    The upper and lower regions differ in their topological phase, so that an extended wave packet travels along the boundary between the regions.
    }
    \label{fig:intro_gyro_symplectic}
\end{figure}

These recent discoveries place a spotlight on the importance of local, real-space physics in determining topological phenomena.
We seek to examine this interplay using gyroscopic metameterials as a concrete foundation for our discussion. 
Our conclusions, however, extend beyond this particular class of systems. 
After revisiting the mathematical foundations of these odd materials, we present new insights into the symmetries underlying these materials' band gaps.
We then adopt a local, mechanical viewpoint on the mechanisms generating topological gaps. 
This extends the current understanding of the mesoscopic or quasi-local character of band topology that has emerged based on studies of gyroscopic materials. 

Lastly, we present experimental and numerical measurements of phase transitions driven by disorder and the localization properties of gyroscopic materials' band structures.
In contrast to recent studies of critical behavior in amorphous topological insulators~\cite{ivaki_criticality_2020,sahlberg_topological_2020}, we find the scaling behavior of critical exponents consistent with the universality class with Cartan label A. 
In addition, a topological Anderson insulator phase diagram shows re-entry into the topological phase with increasing disorder strength, both in experiments and simulations.
\section{An odd interaction and its symplectic dynamics}

Our system is composed of coupled spinning gyroscopes suspended from a plate shown in~\Fig{fig:intro_gyro_symplectic}.
When the spinning speed is large, the high frequency nutation becomes negligible in amplitude.
A gyroscope with a hanging orientation vector $\hat{n}$ subjected to forces acting at $\ell \hat{n}$ exhibits motion in the $xy$ plane governed by the action of torques redirecting the angular momentum axis: 
\begin{align} \label{gyrox}
\frac{I \omega_0}{\ell } \partial_t{x}
& = \ell \frac{\partial U}{\partial y} \\
\label{gyroy}
\frac{I \omega_0}{\ell} \partial_t{y}
& = - \ell \frac{\partial U}{\partial x} ,
\end{align}
where $\ell$ is distance from the pivot point to the point where forces act, $x$ and $y$ are the in-plane displacements of the same point, $U$ is the potential energy of the gyroscope from pinning forces (like gravity and the springs connecting the gyroscope to its neighbors), $\omega_0$ is the spin frequency, $I$ is the moment of inertia along the spinning axis, and $x$ and $y$ are the displacements from the equilibrium position.

\Eqs{gyrox} and~\ref{gyroy} reflect the curious behavior of gyroscopes: they respond to an applied force by moving in a direction transverse to that force. 
Building a material out of coupled gyroscopes translates this unusual response into new collective material excitations that are profoundly different from those of inertial materials, where we are accustomed to having displacements follow the direction of forces.

In conventional elastic materials, waves emerge from the conversion of potential energy into kinetic energy and back again. 
As energy sloshes back and forth between two forms, disturbances propagate through the material. 
For interacting gyroscopes, however, inertia is dominated by the rapid spin of each gyroscope, rendering the kinetic energy of the gyroscopes' center of mass motion irrelevant. 
Instead, the independent degrees of freedom are the orthogonal gyroscope displacement $x$ and $y$, and the resulting energy oscillation is an oscillation of potential energy between these spatial degrees of freedom because of the transverse response of gyroscopes to applied forces.
These dynamics are analogous to those of Tkachenko waves in vortex crystals~\cite{sonin_vortex_1987}.

The relationship between motion in these two directions forms a symplectic structure.
By defining $q = \sqrt{I \omega_0} x / \ell$ and $p = \sqrt{I \omega_0} y / \ell$, we see that~\Eq{gyrox} and~\Eq{gyroy} are the same as Hamilton's equations:
\begin{equation}
\label{eq_hamilton}
\partial_t{q} = \frac{\partial U}{\partial p} 
 \hspace{30pt}
\partial_t{p} = - \frac{\partial U}{\partial q},
\end{equation}
which provides symplectic structure to the dynamics.
\subsection{Symplectic structure of cyclotron orbits}

\begin{figure}
    \centering
    \includegraphics{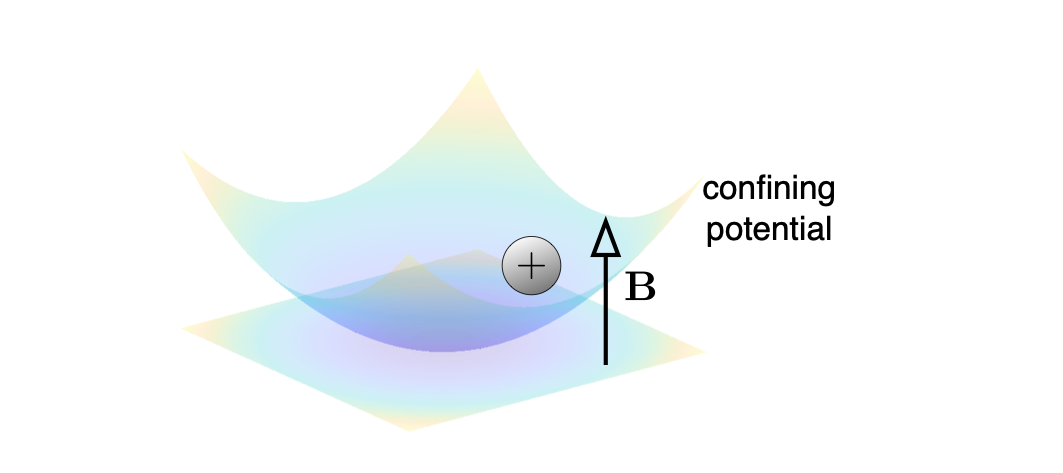}
    \caption{
    \textbf{A charged particle in a magnetic field provides a simple setting in which to understand symplectic dynamics. }
    A particle of positive unit charge moves in a confining potential (blue gradient) and experiences cyclotron orbits due to an out-of-plane magnetic field of strength $B$.}
    \label{fig:charged_particle}
\end{figure}
    
To gain some insight into the practical consequences of these conservative transverse dynamics, it is helpful to examine a simple related system, a charged particle in a magnetic field confined by an external potential (\Fig{fig:charged_particle}).
This simple example highlights subtleties in this category of problems with time-reversal-broken dynamics.

We take the energy to be given by 
\begin{equation} 
H = \frac{1}{2m} \left( \vec{p} - \frac{q\vec{A}}{c} \right)^2 + \frac{k}{2} \left(x^2 + y^2\right),
\end{equation}
where we adopt the gauge such that $\vec{A} = Bx \hat{y}$ for the vector potential $\vec{A}$ so that $\vec{B} = B\hat{z} = \nabla \cross \vec{A}$ and $\vec{p}$ is the canonical momentum.
In order to fully describe the normal modes of the particle's motion, both realspace coordinates $(x,y)$ and the conjugate momenta ($p_x=m\partial_t{x}$, $p_y=m\partial_t{y} + q B x/c$) must be included. \
Hamilton's equations (\Eq{eq_hamilton}) become
\eq
\partial_t \begin{pmatrix}
x \\ y \\ p_x \\ p_y
\end{pmatrix} 
 = \begin{pmatrix}
0 & 0 & \frac{1}{m} & 0 \\
-\frac{qB}{mc} & 0 & 0 & \frac{1}{m} \\
-\frac{q^2B^2}{mc^2}-k & 0 & 0 & \frac{qB}{mc} \\
0 & -k & 0 & 0 \\
\end{pmatrix}
\begin{pmatrix}
x \\ y \\ p_x \\ p_y
\end{pmatrix} .
\qe
This is simply a recasting of Hamilton's equations seen before in~\Eq{eq_hamilton}. 
Similarly to a quantum mechanical Hamiltonian, the dynamical matrix is a linear operator governing the evolution of the system, but the matrix here is not Hermitian. 
Instead, it is symplectic, as we will see. 
The four eigenmodes of the dynamical matrix trace out elliptical orbits, taking the form of 
\begin{align}
\begin{array}{l l c c c}
e_{\pm} &=\frac{1}{\sqrt{2\omega_{\pm} \left(m \alpha_{\pm}^2+\frac{1}{k}\right)}}
\bigg(\alpha_\pm, 
& \frac{-i\omega_{\pm}}{k}, 
& i m \alpha_\pm  \omega_{\pm}, 
& 1 \bigg)
\end{array}
\end{align}
and their complex conjugates, $\xoverline{e}_+$ and $\xoverline{e}_-$,
where the frequencies of the orbits are $\pm \omega_{+}$ and $\pm \omega_{-}$.
Above, 
\begin{equation}
\alpha_{\pm} =
\frac{c}{qBk}\left( k -m \omega_{\pm}^2  \right) 
\end{equation} 
and the eigenmode frequencies are 
\begin{align}
\omega_{\pm} &= \sqrt{\frac{k}{m} + \frac{q^2B^2}{2c^2m^2}\left(1 \pm \sqrt{\frac{4c^2km}{q^2B^2} + 1}\right)}.
\end{align}
These eigenmodes, which correspond physically to cyclotron orbits of the particle precessing in the magnetic field, 
are orthogonal under the symplectic form, defined for any two modes $\psi_\alpha$ and $\psi_\beta$ as 
\eq \label{inner_product}
\langle \psi_\alpha | \cdot^\perp |  \psi_\beta \rangle
= i \sum_{j \in \{1,2\}} 
\xoverline{x_{\alpha,j}} \,  {p_{\beta, j}}
- \xoverline{p_{\alpha,j}}   \, {x_{\beta,j}},
\qe
where $\vec{x} = (x, y)$ and $\vec{p} = (p_x,p_y)$ and $j$ runs over the two spatial dimensions.

This highlights a difference between parity-breaking systems and more common  inertial systems that do not break parity. 
In the latter case, there is a direct proportionality between momentum and velocity.
For an eigenmode in an inertial system, the momentum is then simply proportional to the position offset by a phase ($ p_j = -i m_j \omega x_j$).
This simplifies the dynamical matrix considerably to produce eigenvectors that are orthogonal under both the symplectic form and the familiar Euclidean norm defined as $\sum_j m_j x_{\alpha, j} \xoverline{x_{\beta, j}}$.
This feature persists in systems of many coupled inertial objects. 
In contrast, rotational coupling between motion in $x$ and $y$ prevent the eigenvectors from being orthogonal under the Euclidean inner product.
However, under the symplectic form, the eigenmodes \textit{are} orthogonal for any value of magnetic field $B$ according to
\begin{equation} \label{eq:innerProductPoisson}
\langle {\psi}_{\alpha} |\cdot^\perp | \psi_{\beta} \rangle \propto \delta_{\alpha\beta} \,  \textrm{sign}(\omega_{\alpha}).
\end{equation}

In systems which break parity, the only appropriate inner product is given by~\Eq{inner_product}, and this norm relates the time-averaged energy of a mode, $\langle U \rangle$, to the oscillation frequency, $\omega$.
This relationship is given by 
\begin{equation}\label{eq:energy-frequency}
    \omega_{\alpha} = \frac{\langle U \rangle}
    { \langle \psi_{\alpha} | \cdot^\perp |  \psi_{\alpha} \rangle },
\end{equation}
which holds only when $\langle \psi_{\alpha} | \cdot^\perp |  \psi_{\alpha} \rangle$ is the symplectic form of the eigenmode given in~\Eq{inner_product}.
This simple relationship allows one to estimate oscillation frequencies by having an intuitive picture of energies involved in a mechanical deformation.

The consequences of symplecticity extend beyond a redefinition of the inner product: a symmetry between eigenvectors at positive and negative frequencies appears. 
Each mode appears twice, with one eigenvector representation at frequency $\omega$ and a second, redundant eigenvector with frequency $-\omega$ and conjugated position and momentum, but each pair represents the same physical circular cyclotron orbit. 
This redundancy in the system foreshadows the emergence of ``particle-hole" symmetry in networks of gyroscopes.

\begin{figure}
    \centering
    \includegraphics[width=\linewidth]{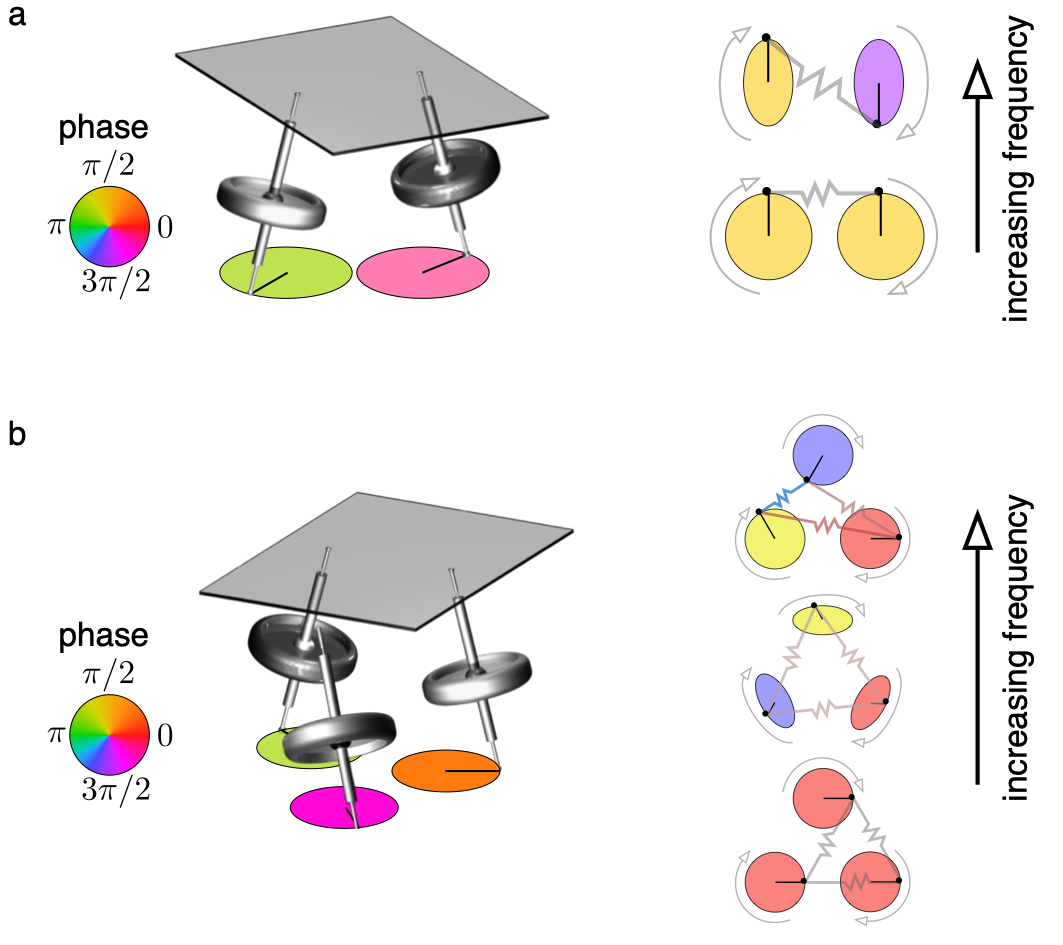}
    \caption{\textbf{Interacting gyroscopes highlight the interplay between clockwise and counter-clockwise polarizations in elliptical eigenmodes.} 
    Eigenmode displacements are represented by ellipses traced out by the gyro tips, each colored by its phase in the ellipse at a snapshot in time.
    \textit{(a)} Two interacting gyroscopes support an in-phase circular motion `ground state' ($\omega=\Omega_p$) and out-of-phase elliptical motion `excited state' ($\omega > \Omega_p$).
    \textit{(b)} Three interacting gyroscopes generate two excited states in addition to an in-phase circular motion `ground state' ($\omega=\Omega_p$).
    One has elliptical motion with relative phases staggered by $2\pi/3$, while the highest frequency motion is again circular, with relative phases staggered by $-2\pi/3$.
    In this figure, we let $\Omega_k=\Omega_p$.
    }
    \label{fig:2gyro}
\end{figure}


\subsection{Coupled gyroscopes yield coupled polarizations}
Gyroscopic metamaterials similarly break parity, mixing momentum and positions and supporting waves that exchange displacements in orthogonal spatial dimensions. 
We consider the motion of the gyroscopes about their hanging rest configuration.
For small displacements $(x, y)$ of the centers of mass, the equation of motion for a gyroscope takes the form
\begin{equation} \label{simulation}
\partial_t
\left( 
\begin{array}{c}
x_i\\
y_i
\end{array}
\right)
 \approx
\Omega_p \left( \begin{array}{c}
y_i  \\
- x_i
 \end{array} \right)
 +
 \frac{\ell^2}{I \omega_0}
\sum_{j=1}^{\textrm{N}}
\left( \begin{array}{c}
-F_{ij, y} \\ F_{ij, x}
\end{array} \right),
\end{equation}
where $\Omega_p$ is the pinning frequency due to gravity, restoring forces from the method of suspension, and other site-specific restoring potentials such as a magnetic field generated by coils placed beneath each gyroscope  (see~\Fig{fig:intro_gyro_symplectic}b-e and~\cite{nash_topological_2015}) and the sum is taken over all $N$ neighbors connected to site $i$.
It is useful to write the displacement of the $p^{th}$ gyroscope as $\psi_p \equiv x_p + i y_p$, so that 
\eq\label{eq:psiRpsiL}
\psi_p = \psi^R_p e^{-i\omega t} + \overline{\psi^L_p} e^{i \omega t} .
\qe
If the frequency of oscillation $\omega$ is positive, the first term encodes clockwise motion, while the second encodes counter-clockwise motion, and $\overline{\psi^L_p}$ is the complex conjugate of $\psi^L_p$.
 
How do these dynamics play out in the simplest scenario of two interacting gyroscopes?
One possible motion is uniform precession at a frequency $\Omega_p$, which does not stretch or compress the bond connecting the two gyros, shown in the lower mode of~\Fig{fig:2gyro}a.
This motion is entirely in-phase. 
The other possible motion is out-of-phase, which does stretch and compress the bond and results in a higher frequency.
This out-of-phase motion is elliptical, not circular, as shown in the upper mode of~\Fig{fig:2gyro}a.
Elliptical traces reflect a superposition of both clockwise and counter-clockwise polarizations, foreshadowing that the interaction between polarizations in the dynamical matrix will play a central role in understanding gyroscopic metamaterials.
As in the case of a charged particle in a magnetic field, here each of these modes is represented twice -- with one eigenstate of frequency $\omega$ dominated by $\psi^R$ and one of frequency $-\omega$ dominated by $\psi^L$.
Physically, both $\psi^R$ at positive frequency and $\psi^L$ at negative frequency represent the same clockwise motion according to~\Eq{eq:psiRpsiL}.

By casting the equations of motion in the basis of $(\psi^R, \psi^L)$ as 
\begin{multline}\label{eq:psi-eom}
i \partial_t \left(
\begin{array}{c}
 \psi^R_p \\
\psi^L_p
\end{array}
\right) 
 = \Omega_p \left( \begin{array}{c}
\psi^R_p \\
-\psi^L_p 
\end{array}
\right)
\\
 + \frac{ \Omega_k}{2}
\sum_{q \in \textrm{N}(p)} \left(
\begin{array}{c}
\psi_{p}^R -\psi_{q}^R 
+ e^{i 2 \theta_{pq}}(\psi_p^L - \psi_q^L) \\
-\psi_{p}^L +\psi_{q}^L 
- e^{-i 2 \theta_{pq}}(\psi_p^R - \psi_q^R ) 
\end{array}
\right),
\end{multline}
where $\Omega_k = k \ell^2/ (I \omega_0)$ is the characteristic spring frequency, $\theta_{pq}$ is the bond angle between gyroscopes $p$ and $q$ fixed by network geometry, and the sum over $q$ is over all of $p$'s coupled neighbors,
we see that clockwise and counter-clockwise polarizations are coupled through the geometry of bonds between gyroscopes. 
Here, $k$ is the spring constant, $\ell$ is the length of the pendulum from the pivot to the point of applied forcing, $I$ is the moment of inertia along the spinning axis, $\omega_0$ is the angular spinning frequency about the long axis of the gyroscope.


Now that we have gained intuition from the two-gyroscope case, we can formulate the many-gyroscope case as an eigenvalue problem.
\Eq{eq:psi-eom} defines the entries for a system's dynamical matrix, $ \mathbf{D}$, such that
\eq \label{eq:basicD}
 i\partial_t \vec{\psi}  = \mathbf{D} \vec{\psi},
\qe
where the components of $\vec{\psi} = (\psi^R, \psi^L)$ encode the displacements of the gyroscopes.
The matrix elements $D_{ij}$ describe the response of gyroscope $j$ to displacements of gyroscope $i$. 


\subsection{Symplecticity affects eigenmodes}

The dynamical matrix is not Hermitian.
However, due to the symplectic symmetry of our dynamical matrix, all eigenvalues of $\mathbf{D}$ are nonetheless real. 
The symplectic symmetry of our dynamical matrix can be stated as:
\eq
D = Q D^\dagger Q,
\qe
with 
\eq
Q = \begin{pmatrix}
\II_n & 0 \\
0 & -\II_n
\end{pmatrix}.
\qe
Just as the real-valued frequencies of the charged particle in a magnetic field come in positive and negative pairs, so too the frequencies of gyroscopic motion arise in positive and negative pairs. 
These two copies of the system's dynamical response are related by a particle-hole symmetry: each eigenmode has two redundant descriptions with frequencies $\pm \omega$ related by conjugation of the gyroscope positions and momenta. 
This is analogous to particle-hole symmetry in electronic systems, wherein the behavior of an electron's excitated state above the Fermi level is tied to the behavior of a electron `hole' left behind below the Fermi level.  
In later sections, we find that the symplectic form must be used in the proper definition of Berry curvature and projector operator elements to build real-space descriptions of band topology. 

\begin{figure}
    \centering
    \includegraphics[width=\linewidth]{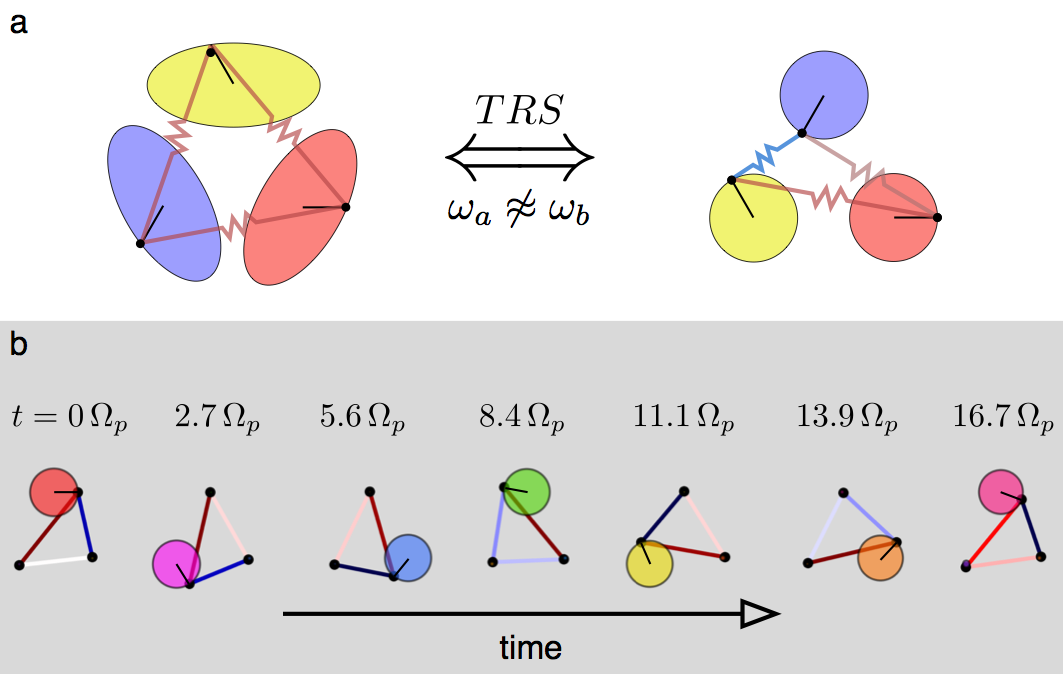}
    \caption{\textbf{Gyroscopic triad provides intuition for the effects of time reversal symmetry breaking.}
    \textit{(a)} Time reversal operation, given by $\psi\rightarrow \bar{\psi}$ and $t\rightarrow -t$, transforms the instantaneous displacements (black bars) of one eigenmode into another for the gyroscopic triad. 
    Here, the instantaneous displacements differ between the two pictured eigenmodes only by $y\rightarrow -y$. 
    Since eigenmodes $\vec{\psi}_{\omega_a}$ (left) and  $\vec{\psi}_{\omega_p}$ (right) have different frequencies, and given that the time reversal operator conjugates these displacements, we see that time reversal symmetry is broken.
    \textit{(b)} Simulating the dynamics of the triad reveals strong signatures of chirality from time reversal symmetry breaking. 
    Because of the phase interference between the different eigenmodes, the displacement generally propagates counter-clockwise between the three gyroscopes. 
    For special values of $\Omega_k/\Omega_p$, the oscillation is periodically localized to one site at regular intervals, and the site of localized excitation advances in a chiral fashion, here shown for $\Omega_k / \Omega_p = 0.4$.
    We note that the excitation is not fully localized at times between the snapshots shown. 
    }
    \label{fig:3gyro_TRS}
\end{figure}

\subsection{Time reversal symmetry breaking}
An interesting transformation to consider is  $\psi\rightarrow\bar{\psi}$ and $t\rightarrow -t$. 
This transformation is analogous to the time reversal symmetry (TRS) operator of the Schr\"odinger equation, which resembles~\Eq{eq:psi-eom}. 
The conjugation operation $\psi \rightarrow \bar{\psi}$ corresponds to reflecting the displacements $y\rightarrow -y$ of each gyroscope about its pivot point, instead of reversing momenta. 
A single gyroscope system clearly respects this symmetry, in contrast to a charged particle in a magnetic field; spinning alone does not break this form of TRS. 
Interestingly, a system of two interacting gyroscopes also respects TRS.

If the bond connecting the two gyroscopes is aligned with the $x$ axis, this follows trivially from the fact that $\theta_{pq} = 0$ or $\pi$. 
Even if the angle $\theta_{pq}$ is nonzero, we can globally rotate the system without penalty.
Therefore, after this coordinate change, the factor $e^{i2\theta_{pq}}$ is purely real, and there is no phase associated with the coupling between polarizations $\psi^R$ and $\psi^L$. 

Introducing a third gyroscope, however, allows time reversal symmetry to be broken.
Now no global rotation of the system will result in bond angles $\theta_{pq}$ such that all factors $e^{i 2 \theta_{pq}}$ are real.
The simplest example in which one can observe the role of TRS breaking is the equilateral triad of coupled gyroscopes shown in~\Fig{fig:2gyro}b. 
For this triad, three distinct eigenmodes arise: one uniform in-phase precession with no spring stretching, and two `excited state' motions $\vec{e}_a$ and $\vec{e}_b$. 
In the particular case of the equilateral triad, neighboring gyroscopes differ in phases by $2\pi/3$ either clockwise (for $\vec{e}_a$) or counterclockwise (for $\vec{e}_b$) around the triangle. 
The action of the time reversal operation mixes eigenstates. 
This can be seen intuitively by examination of the excited states shown in~\Fig{fig:3gyro_TRS}a.
In particular, the $TR$ operation, given by $\psi \rightarrow \bar{\psi}$ and $t\rightarrow -t$, mixes $\vec{e}_a$ and $\vec{e}_b$, which have different frequencies.
In this scenario, we find that TRS breaking is manifest by the difference in frequency of the two eigenmodes that are mixed by the \textit{TR} operation. 
More generally, for systems of more than two gyroscopes, \textit{TR} transforms a given configuration into a linear superposition of modes with different frequencies unless there are additional symmetries that guarantee the bond energy is preserved: $TR$ is not typically a symmetry of the dynamical equations~\cite{mitchell_tunable_2018}.

This lack of TRS is manifest in the differing frequencies of these eigenmodes. 
Together with the fact that the two excited states have differing relative phases between gyroscopes ($\pm 2\pi/3$ radians), the frequency difference between these modes leads to beating patterns with a chiral flow of energy.
As shown in~\Fig{fig:3gyro_TRS}b, displacing a single gyroscope leads to large amplitude excitation that sloshes from one gyroscope to another in a chiral fashion over time.
This resulting chiral conductance of energy foreshadows the emergence of chiral edge waves.


Now that we have gained intuition for the role of TRS breaking from the three-gyroscope case, we can see the effects of TRS breaking in the dynamical matrix itself. 
We can envision excitation of gyroscopes `hopping' from site to site with couplings given directly by the elements of the dynamical matrix $\mathbf{D}$ in~\Eq{eq:basicD}. 
In fact, we see directly from~\Eq{eq:psi-eom} that the coupling of clockwise precession and counter-clockwise precession depends explicitly on the geometry with \textit{complex} couplings, whereas gyroscopic motion with the same polarization at different sites -- $\psi^R_i$ and $\psi^R_j$, for example -- have purely real couplings.

\begin{figure}
    \centering
    \includegraphics[width=\linewidth]{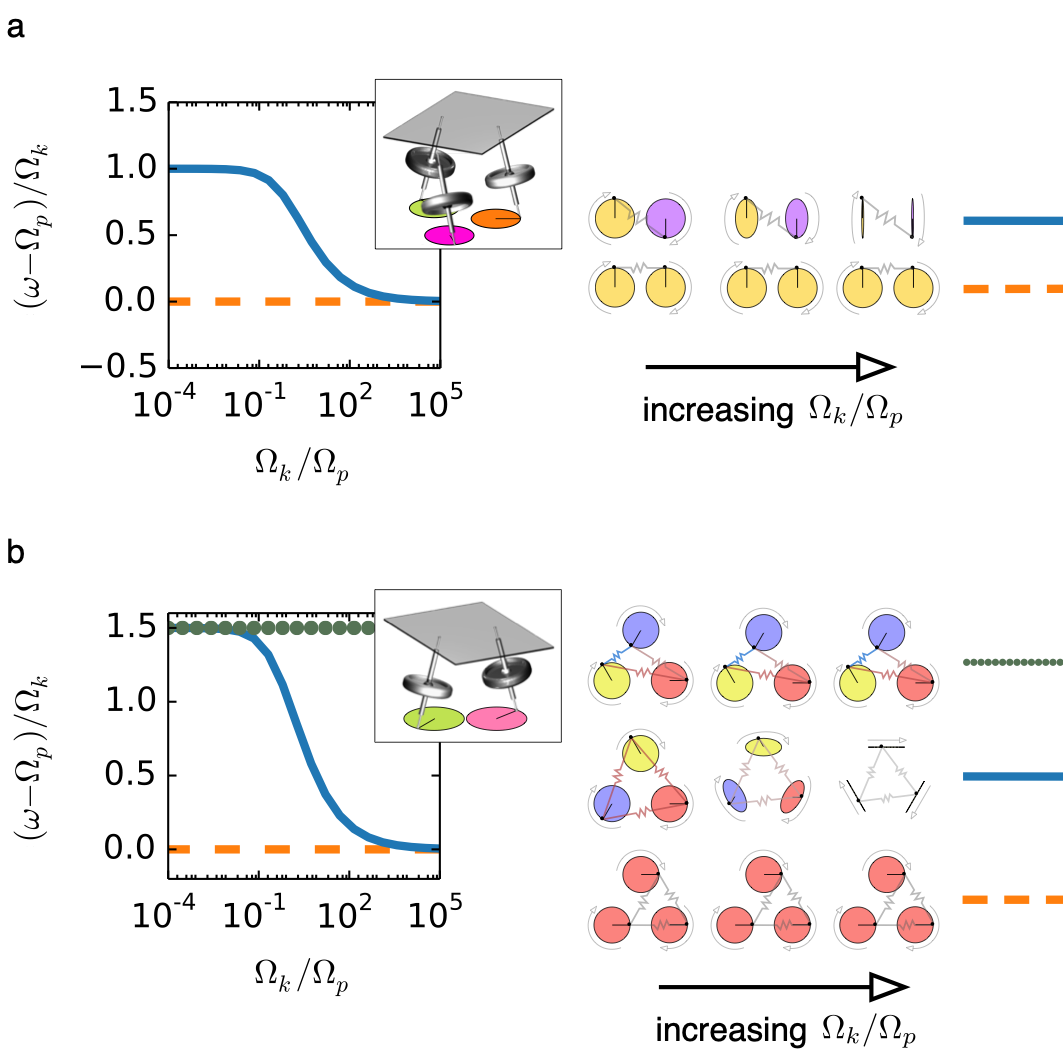}
    \caption{\textbf{Low interaction strength enables connection to tight-binding models by diminishing coupling between polarizations.}
    In the limit that $\Omega_k /\Omega_p \rightarrow 0$, eigenmode frequencies of a gyroscope pair or triad approach the precession frequency, $\Omega_p$. 
    The slight deviations from $\Omega_p$ are proportional to $\Omega_k$, resembling a tight-binding model with inter-site coupling of $\Omega_k/2$ (\Eq{eq:tightbinding_limit_gyro}). 
    \textit{(a)} Two modes represent a pair of gyroscopes' bonding and anti-bonding behavior.
    Decreasing the interaction strength decreases the ellipticity of the higher-frequency mode: the counter-rotating component presents only a small perturbation on the dominant clockwise motion.
    In the limit of $\Omega_k / \Omega_p \rightarrow 0$, both modes are nearly circular, like in quantum mechanical models with unitary evolution. 
    Here, the phase of the displacement is akin to the phase of a wavefunction. 
    \textit{(b)} For the gyroscopic triad, the oscillation frequency is $\Omega_p$ for the uniform precession (`ground state') mode and $\Omega_p + 3\Omega_k/2$ for the phased circular mode. 
    The intermediate mode frequency varies as a function of $\Omega_k/\Omega_p$, interpolating between the two bounds.
    Only the intermediate eigenmode changes shape, becoming increasingly circular as the coupling strength weakens.
    } \label{fig:pinning_limit}
\end{figure}

\subsection{Connection to tight-binding models: the limit of weak interactions}

If we consider the dynamics when interactions between gyroscopes are weak compared to the pinning forces restoring a gryoscope toward its hanging position, the effects of counter-clockwise motion act as a perturbation on the dominant clockwise precession.
\Fig{fig:pinning_limit} highlights how the two-gyro and three-gyro cases explored thus far vary as the interaction strength is tuned.
Modes that mix polarizations become increasingly elliptical as the coupling strength grows.
For weak interactions ($\Omega_k/\Omega_p \ll 1$), gyroscopic networks' eigenmodes become nearly circular, which highlights the perturbative effect of the counter-rotating polarization on the dominant precession.

In fact, in this limit there is a natural correspondence with hopping models ubiquitous in condensed matter.
Just as precession of a gyroscope can exert torque on its neighbors to induce precessional motion, with an energy cost depending on the bond strength, so too electrons or other excitations can hop from one site to another associated energy  based on the coupling between sites.
In this `tight binding limit', in which $\psi^L$ has a perturbative effect on the dominant $\psi^R$ precession, we again see that all complex terms arise from the network geometry, meaning that \textit{phase offsets} are introduced in the response of one gyroscope to another's displacement in a manner that depends on angles between adjacent bonds in the network~\cite{nash_topological_2015}. 
In particular, the approximate equations of motion allow one to predict the evolution by considering only the magnitudes and phases of the dominant rotation at each site:
\begin{multline}\label{eq:tightbinding_limit_gyro}
\omega \psi^R_i = \Omega_p \psi^R_i + \frac{\Omega_k}{2} \sum_{m \in N(i)} (\psi^R_i - \psi^R_j) \\
 + \frac{\Omega_k^2}{8 \Omega_p}\left[ 
 - \sum_{j,j'}\left(\psi^R_i - \psi^R_j\right) e^{2i \theta_{j'ij}} \right.\\
 \left.
 + \sum_{j,k} (\psi^R_j-\psi^R_k)e^{2i\theta_{ijk}}
\right],
\end{multline}
where the bond angle $\theta_{ijk}$ is the opening angle between bonds $ij$ and $jk$, the first sum is over the $N$ neighbors of $i$, and the subsequent sums are over pairs of neighbors $j$ and $j'$ of $i$ and neighbors $k$ of $j$ for each of $i$'s neighbors $j$.  
Note that the phase offsets of $2\theta$ arise in the term proportional to $\Omega_k^2 /8 \Omega_p$.
This complex next-nearest-neighbor coupling is reminiscent of phases from magnetic flux in common tight-binding models~\cite{haldane_model_1988}, but here the phases depend on the arrangement of the gyroscopes in space.
Furthermore, the phase shifts arise here because of the counter-rotating polarizations, $\psi^L$, which are not present in the tight-binding picture.
We stress that the complex-valued coupling encodes the fact that perturbing one gyroscope gives rise to a \textit{phase-shifted} response at a nearby site, with a phase shift dictated by the geometric arrangement of the bonds.
Thus, geometry provides phase shifts which generically break TRS except in special circumstances such as a rectangular lattice, where all angles $\theta_{ijk}$ are multiples of $\pi/2$. 
As discussed below, it is these TRS-breaking terms that generate topological bandgaps in gyroscopic networks.

\section{Band gaps without full diagonalization}

\begin{figure*}[ht]
\includegraphics[width=\textwidth]{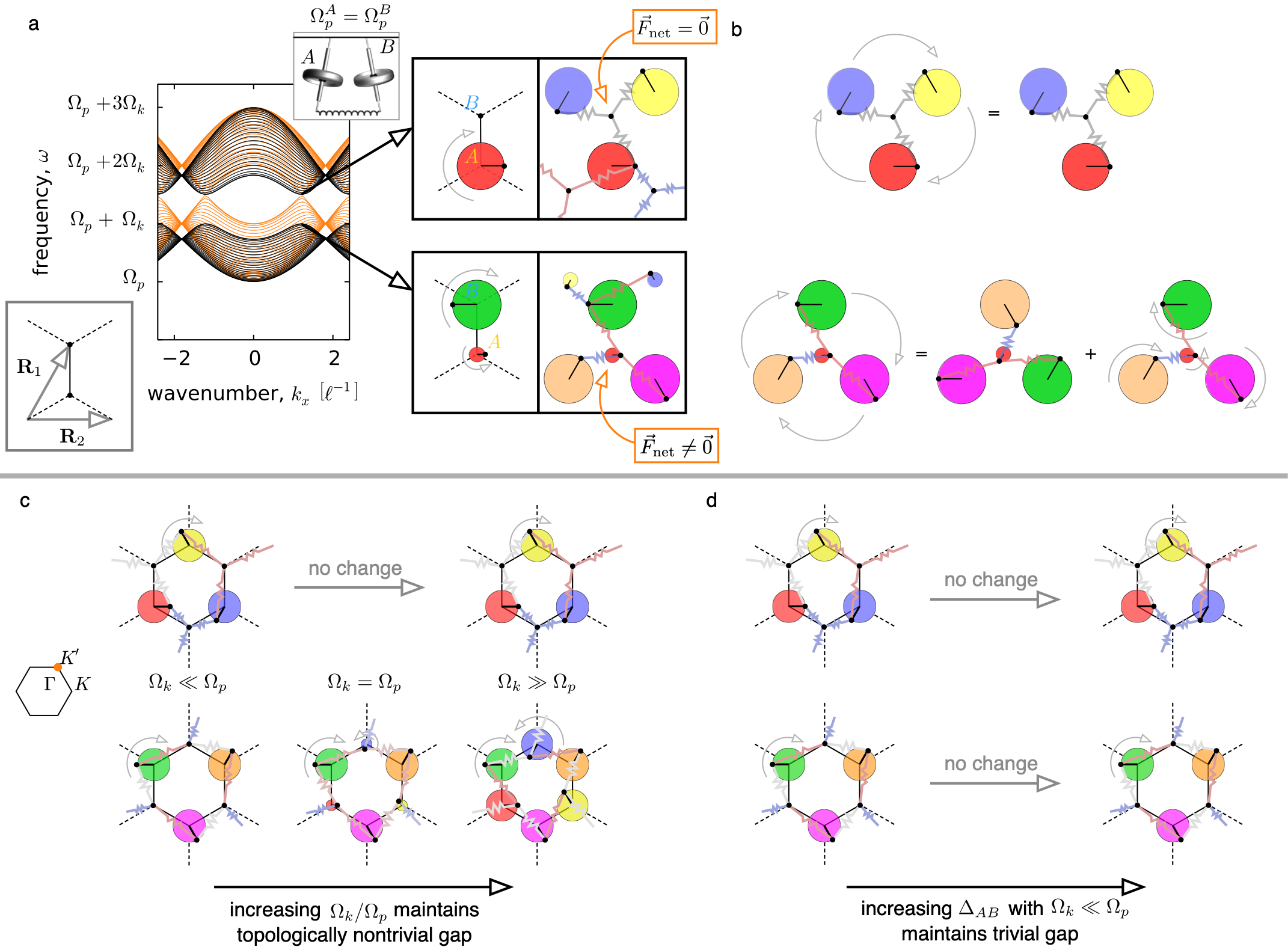}
\caption[]{
\textbf{Gyroscope mode displacements reveal how coupled circular polarizations lead to a gap.}
\textit{(a)}
The Dirac cones for the weak-coupling band structure (light orange bands) evolve into a band gap for the finite coupling case (black band structure), here shown for $\Omega_k = \Omega_p$.
The point of closest approach lies at the wavevectors $\vec{k} = K$ or $K'$, where the normal mode displacements take on a simple character:  clockwise precession of one sublattice site and small (lower) or vanishing (upper) counter-clockwise motion at the second sublattice site (left inset).
Upon examination of the spring couplings connecting a unit cell to its neighbors (right insets), we find that spring forces cancel for the upper mode but do not cancel at the sublattice site with small excitation for the lower mode.
\textit{(b)} The geometry of these eigenmodes is determined by rotational symmetry at the $K$ and $K'$ points. 
Since a $120^\circ$ rotation about any gyroscope must change the pattern only by a phase, all displacements must be purely circular. 
Furthermore, the state can either be invariant under rotation (upper panel) or else phase shifted with a counter-clockwise rotation at the $A$ site (lower panel). 
\textit{(c)} For $\Omega_k \ll \Omega_p$, the two normal modes are asymptotically equivalent, up to an inversion of $A\leftrightarrow B$, since the smaller counter-clockwise rotation on $A$ vanishes in this limit. 
As the spring coupling increases, the counter-clockwise excitation at the $A$ site grows in magnitude, decreasing the spring energy and broadening the gap. In the limit of $\Omega_k \gg \Omega_p$ (rightmost images), the counter-clockwise excitation magnitude approaches that of the clockwise excitation.
\textit{(d)}
When the pinning frequencies at the two sublattice sites differ, so that $\Omega_p^A \rightarrow \Omega_p + \Delta_{AB}$ and $\Omega_p^B \rightarrow \Omega_p - \Delta_{AB}$, a trivial gap opens in the limit of $\Omega_k \ll \Omega_p$. 
Though the lower frequency state does not exhibit torque balance at the immobile sublattice site, the spring energy is negligible compared to $\Omega_p$, so pinning torques dominate and no counter-clockwise motion is observed.
As a consequence, the excitation patterns bounding the gap do not change as the frequency splitting between the $A$ and $B$ sites grows.
}
\label{fig_diracmodes}
\end{figure*}

In this section we consider the spectral gaps that arise naturally in large collections of gyroscopes in ordered as well as amorphous configurations. 
We focus on the role of symmetry and local dynamics in determining their origin.
We begin by examining the mechanical motions that underlie the frequency splitting of neighboring bands, linking these motions to TRS breaking through the coupling of left- and right-circular polarizations.
We then study the evolution of the band structure as we tune the coupling between neighboring unit cells of gyroscope pairs, allowing us to contrast the topological gap origins with gaps created from bonding/anti-bonding behavior of isolated dimers.
Lastly, we provide a simple method to predict the existence of band gaps without full diagonalization of a system's dynamical matrix.

\subsection{Band gaps in lattices}

A simple honeycomb lattice of gyroscopes has gaps in the system's bulk phonon spectrum, shown in the black band structure in~\Fig{fig_diracmodes}a.
If the spring-like coupling between sites is weak, the gap nearly closes at the corners of the Brillouin zone (orange spectrum in~\Fig{fig_diracmodes}a). 
As we increase the coupling, the gap broadens.
Physically, gaps arise due to the coupling of left-polarized motion with the right-polarized motion of the gyroscopes, which breaks time reversal symmetry in the dynamics (\Eq{eq:psi-eom}).
Inspecting the honeycomb lattice's normal modes bounding the gap leads to an intuitive picture of this coupling's effect.

The lattice symmetry allows us to focus on a single hexagonal plaquette to fully describe the bulk excitations bounding the bandgap. 
\Fig{fig_diracmodes}a shows these modes bounding the gap, located at the corners of the Brillouin zone $K$ and $K'$. 
In the higher frequency mode bounding the gap, one sublattice site precesses clockwise and the other does not move. 
In the lower frequency mode, on the other hand, the excitation is not completely localized to one of the two lattice sites: the other site precess counterclockwise with a small amplitude. 
The difference in this motion provides an intuitive way to explain the gap, as we detail in a multi-step argument in~\Fig{fig_diracmodes} and Appendix~\ref{appendix:dirac_splitting}.

The crux of the argument relies on the relationship between energy and frequency given earlier in~\Eq{eq:energy-frequency}. 
The small counterclockwise displacement in the lower-frequency mode relaxes the cycle-averaged spring energy, and since the oscillation frequency $\omega$ is directly proportional to the average potential energy over a normal mode cycle, the frequency of this mode is smaller than the state with a truly immobile site
(see Appendix~\ref{appendix:dirac_splitting} for details). 
This lower energy is evident from the force imbalance from the neighboring springs in the lower-frequency mode (\Fig{fig_diracmodes}a).
In contrast, for the higher-frequency mode, springs are stretched or compressed by the same amount around the $B$ site, so there is no net force, and the center gyroscope does not move at all.
This mechanical mechanism for gap generation is different than in the Haldane model~\cite{haldane_model_1988}, where all excitation resides on only one of the two sublattice sites for both modes bounding the gap and the energy splitting arises purely from the next-nearest neighbor coupling.

In this case, the particular lattice symmetry of the honeycomb allowed us to make strong constraints on the shape of the mode displacements to explain the band gap (see Appendix~\ref{appendix:dirac_splitting}), but this is not always possible in general.
In later sections, in discussing structures without translational order, we will provide alternative arguments for the existence of topological band gaps.
Increasing the spring coupling amplifies the frequency separation between the two states, broadening the gap and leading to larger and larger counter-clockwise rotation on the $B$ sites at $K'$, as shown in~\Fig{fig_diracmodes}c.
If, instead of increasing the spring coupling, we detune the precession frequencies at the two sublattice sites, the gap similarly broadens. 
However, the material is a trivial insulator, and the motions of the two states bounding the gap remain unchanged as the inversion symmetry breaking grows, without any signature of coupling between right and left polarizations (\Fig{fig_diracmodes}d).


What have we learned? 
The interplay between coupled polarizations and gyroscope phases interact in tandem to break time reversal symmetry, reminiscent of the simple three-gyroscope triad of Figures~\ref{fig:2gyro}-~\ref{fig:pinning_limit}.
The spring forces directly induce counter-rotation because of the surrounding gyroscopes' phases.
In turn, this coupling between the clockwise and counter-clockwise states opens the gap by splitting the cycle-averaged spring energies, imparting an opposite chirality of precession between the two sublattice sites. 
In more general scenarios, we may not have such symmetric configurations near the gap, but the role of coupled polarizations and relative phases remains.

Importantly, if the dynamics were artificially modified to suppress the coupling between the left polarized (dominant for $\omega < 0$) and right polarized (dominant for $\omega > 0$) excitations, this splitting would not be possible, and no gap would emerge.
That is, if we adjust the strength of block off-diagonal entries of the dynamical matrix that describe the coupling between clockwise and counter-clockwise motion,
\begin{equation}
    D = \left( 
    \begin{array}{c|c}
    D_{LL} & \varepsilon D_{LR} \\
\hline
    \varepsilon  D_{RL} 
    & D_{RR}
    \end{array}
    \right),
\end{equation}
so that $\varepsilon \rightarrow 0$, then the topological character of the bands vanishes. 
From~\Eq{eq:psi-eom}, we recognize that this modification destroys any dependence of the dynamics on the geometry of the lattice, further highlighting the geometry-dependent topological phases that emerge. 
This right-left-coupling avenue for gap formation is distinct from inversion symmetry breaking, which instead adds only diagonal offsets to the dynamical matrix in $D_{LL}$ and $D_{RR}$, with no mixing of polarizations.

\subsection{Spectrum flow: from isolated fragments to a strongly-coupled network }

\begin{figure*}[ht]
\includegraphics[width=\linewidth]{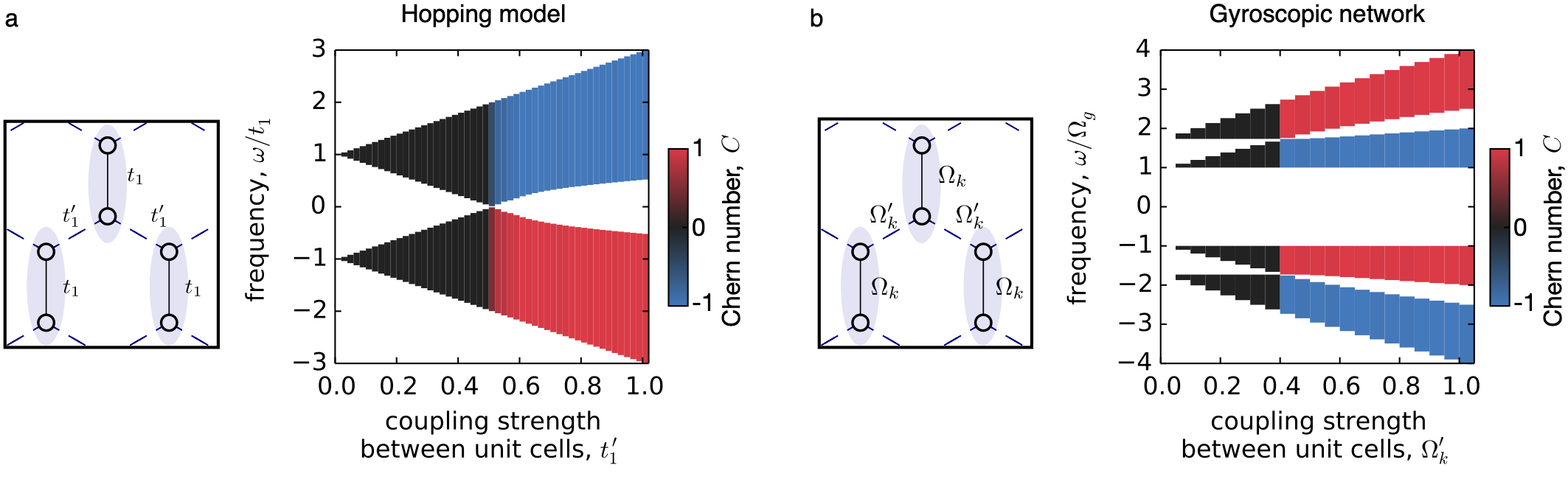}
\caption[]{
\textbf{
The transition from isolated bonding/anti-bonding behavior to Chern insulating behavior in a Haldane-like tight-binding model mimics the analogous transition in gyroscopes.
}
\textit{(a)} Pairs of bonded sites or \textit{(b)} gyroscopes are coupled to each other through  bonds of varying strength (dashed lines). 
When the coupling is absent ($t_1'=0$ or $\Omega_k'=0$), the material is in a trivial insulating phase with one `bonding state', in which the excitations at site pairs are in phase, and one `anti-bonding state', in which the excitations are out of phase. 
These bonding and anti-bonding behaviors thicken into bands as the coupling is increased, until the two mix at the gap closing.
Beyond this value, each band becomes topologically nontrivial, as marked by their nonzero Chern numbers $C = \pm 1$ (red and blue, respectively).
The strength of complex next-nearest neighbor couplings, $t_2$, are controlled by the weakest link between next-nearest neighbors, which is $t_1'$. 
Here, we set $t_2 = i t_1' / 10$. 
\textit{(b)}
The gyroscopic case, governed by the equations of motion given in~\Eq{eq:psi-eom}, mimics the tight-binding picture, though there is a redundant copy of the positive frequency band pairs at negative frequencies.
 }
\label{fig:spectrumflow}
\end{figure*}

One way to disentangle local and global effects in an ordered system is to artificially partition an ordered system into subsections and study what happens as the partitions are coupled.
We do this here with our gyroscopic lattice, showing that the topological gap cannot be understood as frequency splitting between bonding and anti-bonding behaviors, and we find analogous behaviour in a familiar tight binding model.

A pair of coupled oscillators will typically have a frequency splitting between bonding (in-phase) and anti-bonding (out-of-phase) behavior.
Even in the presence of coupling between many such pairs of sites -- which endows width to bands of eigenmode frequencies -- this fundamental origin of the two bands can be seen in the in-phase and out-of-phase character of eigenmodes.  
This results in a local origin of band gaps in tight binding systems, with a natural extension to the case of amorphous materials~\cite{weaire_electronic_1971}. 

It is tempting to think that the two bands of our topological metamaterial seen in~\Fig{fig_diracmodes} may be continuously connected to  in-phase bonding and out-of-phase anti-bonding behavior of gyroscopes in the unit cell.
This, however, cannot be the case. 
We show this in \Fig{fig:spectrumflow} by considering a periodic honeycomb lattice with tunable coupling between unit cells, each of which consists of a pair of sites.
We compare both the tight-binding and gyroscopic case, where the two-banded tight-binding dynamics analogous to~\Eq{eq:tightbinding_limit_gyro} is given by the Haldane model of Ref.~\cite{haldane_model_1988}:
\begin{equation}
H = -t_1 \sum_{\langle ij \rangle} c_i^\dagger c_j - t_2
\sum_{\langle\langle ij \rangle\rangle}e^{-i \phi_{ij}} c_i^\dagger c_j,
\end{equation}
where $\langle ij \rangle$ denotes nearest neighbors, $\langle\langle ij \rangle \rangle$ denotes next-nearest neighbors, $c^\dagger_i$ is the creation operator at site $i$, and $\phi_{ij}=-\phi_{ji}$ is a phase shift associated with hopping from site $i$ to site $j$.
By comparison with~\Eq{eq:tightbinding_limit_gyro}, we see that $t_1$ is akin to the coupling $\Omega_k$ between gyroscopes, $t_2$ is akin to $\Omega_k^2/8\Omega_p$, and $\phi_{ij}$ plays the role of the gyroscopic network bond angle $2\theta_{ijk}$. 
To tune between bonding/anti-bonding behavior within each pair of sites and topological behavior across the bulk, we vary the direct coupling $t'_1$ between adjacent unit cells while keeping $t_1$ fixed for sites within a unit cell, and we set $t_2=i t_1'/10$.

If the bonds denoted by dashed lines are severed completely ($t_1' = 0$), the resulting network consists of isolated pairs of sites, each coupled by a bond of strength $t_1$ and uncoupled from other pairs. 
Such a system is a trivial insulator, akin to isolated molecular dimers in a lattice. 
As the dashed bonds are strengthened, the width of the two bands begin to broaden while remaining in the trivial insulator phase.
The system transitions to a topological phase once $t'_1 \approx 0.5$.
Inspection of the eigenstates near the topological gap for $t_1' > 0.5$ reveals that the displacement patterns do not resemble simple bonding/anti-bonding patterns (refer back to the normal modes for $\Omega_k \ll \Omega_p$ in ~\Fig{fig_diracmodes}c, which correspond to the case of $t_1'=t_1$). 
How do we interpret this? 
For small $t'_1$, the gap is formed by a frequency separation between bonding and anti-bonding behavior, while the gap with large $t'_1$ arises due to the phase shifts enabled by coupling between clockwise and counter-clockwise polarizations. 
Furthermore, the two phases have a macroscopic difference: the topological phase supports chiral edgemodes, indicated by the nonzero Chern number of each band.
In~\Fig{fig:spectrumflow}, the colored bands exhibit nonzero topological invariants of $C=+1$ (red) and $-1$ (blue), which we will define in depth in section~\ref{section:real-space_topology}.
In short, by interpolating between the trivial and topological phases, we find that bonding and anti-bonding bands mix in order to open a topologically nontrivial gap.


The gyroscopic case is similar: trivial bonding/anti-bonding gaps are not continuously connected to the topological gaps from TRS breaking.
As shown in Figure~\Fig{fig:spectrumflow}b, the spectrum of the gyroscopic honeycomb lattice with spring interactions resembles two copies of the Haldane model spectrum, one with $\omega > 0$ and another with $\omega < 0$, with symplectic symmetry relating the two copies. 
(Elsewhere in the text, we omit the redundant, negative frequency states.)
Varying $\Omega'_k$ between adjacent unit cells drives a topological phase transition, closing the gap generated by bonding/anti-bonding behavior of the coupled dimers. 
Once the gap reopens, the bonding and anti-bonding behaviors are mixed together, allowing topologically non-trivial bands and chiral edge modes in the gaps.

In this strongly interacting regime, the classic argument for the local origin of bandgaps (like that of~\cite{weaire_electronic_1971}) does not apply.
This is because the physics of TRS breaking that opens the gap is distinct from the bonding/anti-bonding picture.
The question is then: can we still predict whether there is a gap using a local argument, or do we have to  compute the full band structure?
\section{Understanding the gap in the strongly interacting regime}
\label{section:understanding_gap_in_regime}


The origin of band gaps can be easily understood in systems in which individual sites are weakly coupled to each other, as compared to their on-site energy. 
The dynamical matrix of such a system has the general form of $D=D_{diag}+\delta$,
where $D_{diag}$ encodes the on-site energy in the electronic tight binding picture (and the precession frequency of the individual gyroscopes in the gyroscopic case). 
$\delta$ is a matrix of weak (as compared to the elements of $D_{diag}$) off-diagonal couplings. 
Schematically, 
\begin{equation}\label{ex_staggered}
D \approx 
\left(\,\,
\begin{array}{ccc}
    \diagentry{\begin{array}{cc} t_0 & \\& -t_0\end{array}} & & \delta \\
    &\diagentry{\xddots}\\
    \delta &&\diagentry{\begin{array}{cc} t_0 & \\ & -t_0\end{array}}
\end{array} \, \,
\right).
\end{equation}
For $\delta \ll |t_0|$, the eigenvalues of $D$ are simply close to the diagonal entries, and therefore gaps reside between different values of diagonal elements.
In the example of~\Eq{ex_staggered}, the eigenvalues bunch together around $\pm t_0$, and a spectral gap lies in between. 

For trivial insulators with bonding/anti-bonding behavior, the dynamical matrix is not nearly diagonal, but can be arranged to be nearly block diagonal.
The dynamical matrix of such a system appears in nearly block diagonal form, with all strong couplings contained within blocks and only small terms in the off-diagonal sectors:
\eq 
D \approx 
\begin{pmatrix}
    \diagentry{\begin{array}{cc} 0& t\\t& 0\end{array}} & & \delta \\
    &\diagentry{\xddots}\\
    \delta &&\diagentry{\begin{array}{cc} 0& t\\
    t& 0\end{array}}\\
\end{pmatrix}.
\qe 
As highlighted in the previous section, a system that can be divided into pieces, like our weakly coupled unit cells in a honeycomb lattice, generates two separated frequency bands from bonding and anti-bonding.
The weak connections between the components broaden the spectrum by only a small amount, leaving the gap intact.
This is similar to previous works in which a bond-centric description can explain why amorphous glasses have band gaps~\cite{weaire_electronic_1971}.
Note that in this case, though $D$ is not nearly diagonal, $D^2$ \textit{is} nearly diagonal. 

Unlike either diagonal or block diagonal systems, a system with a topological band gap cannot be divided up into nearly independent pieces (\Fig{fig:spectrumflow}, see also~\cite{arovas_localization_1988}). 
To transform a trivial gap into a topological one, couplings must be sufficiently strong to close and reopen the gap. 
From the perspective of the dynamical matrix, these strong off-diagonal couplings cannot be arranged in block diagonal form~\cite{arovas_localization_1988,huo_current_1992}. 
For example, the structure of such a dynamical matrix could follow the structure of~\Eq{eq:psi-eom}, which we can roughly conceptualize as 
\begin{equation}
D \approx 
\begin{pmatrix}
    \diagentry{0} & ...  & t & ... & it' \\
    &&\diagentry{\xddots}& &\\
    -it' & ... & t & ... &\diagentry{0}\\
\end{pmatrix},
\end{equation}
where large couplings $t$ and TRS breaking terms $i t'$ are distributed throughout. 
One can certainly characterize band gaps by diagonalizing the Hamiltonian, but it would be useful to have a protocol for estimating whether a gap exists or what frequencies a gap may occupy without this costly step.
For the case of a broad class of topological systems, we note that although $D$ is highly non-local, $D^2$ is nearly diagonal, and we construct a method for estimating the gap in these systems below. 
This condition is closely connected to the condition of having a flat band, as occurs in crystal materials supporting the Fractional Quantum Hall Effect~\cite{leykam_artificial_2018}. 
Such a protocol lends itself particularly well for understanding gaps in amorphous systems.

\subsection{Existence of a band gap: algebraic approach}








Consider a system with a dynamical matrix $D$ such that (1) some off-diagonal elements of $D$ are large compared to on-diagonal elements -- ie, there are some couplings that are large compared to on-site pinning -- and (2) the square of the dynamical matrix is nearly diagonal such that
\begin{equation} \label{eq:interference}
D^2=\omega_0^2\mathds{1}+\epsilon,
\end{equation}
where $\omega_0$ is defined as the square-root of the diagonal entry of $D^2$ and $\epsilon$ contains all the off-diagonal entries.
We seek an argument that indicates whether a band gap exists without diagonalization of $D$.

 \begin{figure}
 \includegraphics[width=.45\textwidth]{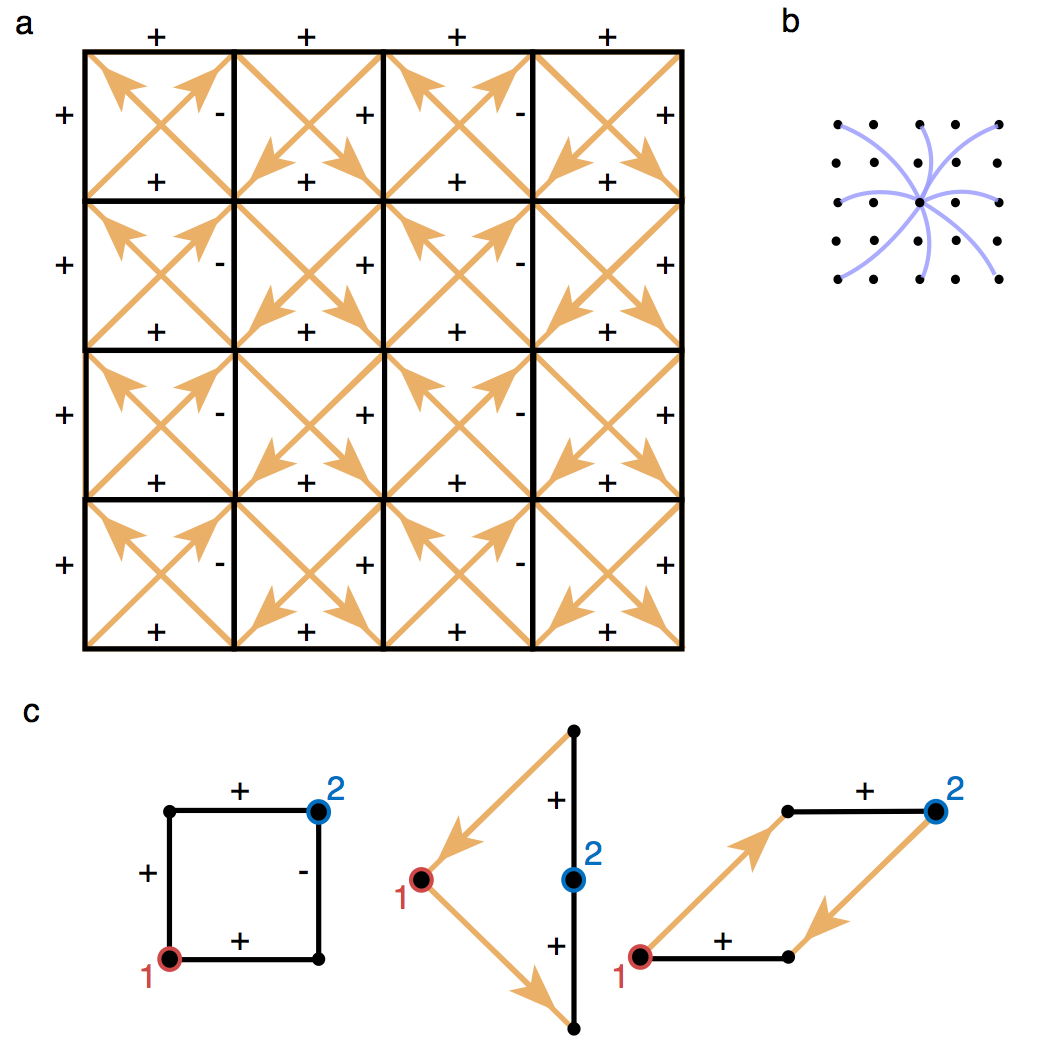}
 \caption{Square lattice model whose band gap and Chern number can be calculated by a local approximation. 
\textit{(a)} The model's dynamical matrix $D$ is built from couplings $\pm t$ (solid lines) with the sign drawn on the picture.  
 The grey diagonal lines have a coupling $i t'$ in the direction indicated by the arrows and $-it'$ in the opposite direction. 
\textit{(b)} The square of the dynamical matrix, $D^2$, includes only horizontal and vertical couplings of two steps with matrix-entry $t^2-2t'^{2}$ and diagonal couplings with matrix entry $t'^2$.  Because of the phases in $t$, all the other matrix elements cancel out.  
 \textit{(c)} Examples of pairs of paths that cancel one another in calculating $D^2$ for the matrix element connecting site 1 (red) to site 2 (blue).}
  \label{fig:Hofstadter}
 \end{figure}
 
Both conventional insulators with strong bonding/anti-bonding behavior and topological systems meet these criteria, provided one subtracts a constant frequency so that the gap is approximately centered about zero.
As an example with which to make the argument concrete, we will use one of the simplest systems with a Chern number, shown in~\Fig{fig:Hofstadter}.
  This model comes from studying electrons in a square lattice in a magnetic field (a Hofstadter model) with a half unit of magnetic flux per unit cell. 
  The diagonal couplings of strength $t'$ break time reversal symmetry, opening the possibility of a topological band gap, while direct couplings are given a strength $t$.
We choose signs and phases of the couplings so that the phase of all the couplings going around each right triangle is $\pi / 2$.  
Using this  model, we find that the gap can be understood to result from interference during propagation of waves in the system, and a variational principle shows that this system has a gap for a range of values of the complex, next-nearest-neighbor coupling strength,  $t'/t$. 
The method presented below can also be applied to a system of gyroscopes in the weak-interaction limit as described by~\Eq{eq:tightbinding_limit_gyro}.


We first show that there are both positive and negative
eigenvalues (i.e., eigenvalues both above and below the net pinning frequency of an individual gyroscope). 
The value of $v^\dagger D v$ for any normalized vector $v$ is an upper bound for the lowest eigenvalue of $D$ if $D$ is Hermitian, according to the variational principle.  
Similarly, for another vector $u$, $u^\dagger D u$ is a lower bound on the highest eigenvalue. 
One can choose in a variety of ways vectors $v$ and $u$ such that $v^\dagger D v$ is either positive or negative, which forces there to be at least some positive and some negative eigenvalues. 
For example, let $v$ be non-zero only on two adjacent sites of the network; the expectation value is either positive or negative for a suitable choice of relative phase between the two sites.
This will be the case if there are off-diagonal elements that are larger than the diagonal elements of $D$ -- which is certainly true for the Hofstadter model, for which diagonal elements are zero.

Not only are there states above and below, but furthermore these two collections of states do not touch. 
To see this, consider the case where all diagonal couplings in the dynamical matrix are equal, and note that the \textit{square} of dynamical matrix has off-diagonal couplings that are small relative to diagonal ones: $D^2=\omega_0^2\mathds{1}+\epsilon$. 
If the entries of $\epsilon$ are small compared to $\omega_0^2$ the eigenvalues of $D^2$ must be close to $\omega_0^2$, so $D$'s eigenvalues must be close to $\pm \omega_0$, leaving a gap near $0$.  
To make this argument more precise, we use a bound from Appendix~\ref{appendix:proof_gap}.
From~\Eq{eq:interference} it follows that the eigenvalues of $D^2$ are greater than or equal to $\omega_0^2-\max_i \sum_j|\epsilon_{ij}|$. 
If this is positive, $D$ has a gap whose width $\Delta$ is at least
 \begin{equation}
 \Delta \ge 2\sqrt{\omega_0^2-\max_i\sum_j|\epsilon_{ij}|}.
 \end{equation} 
 Thus, the width of the gap is bounded by a quantity that does not require the diagonalization of $D$.

Intuitively, we can see that $D^2$ is nearly diagonal because of the effects of interference between paths.
 Each entry of $D^2$ is a sum $\sum_k D_{ik}D_{kj}$ over intermediate vertices $k$ that have a nonzero coupling to both vertex $i$ and vertex $j$.
 For the $D$ given above, the entries of $D^2$ are shown in \Fig{fig:Hofstadter}b.  
 Any pair of sites that are connected by a sequence of two bonds can in principle have a nonzero matrix element, but many of the couplings are zero because they cancel out (\Fig{fig:Hofstadter}c).
 The quantities in the estimate are $\omega_0=2\sqrt{t^2+t^{'2}}$ and $\max_i\sum_j|\epsilon_{ij}|=4|t^2-2t^{'2}|+4t^{'2} $ so $\Delta \ge 4\sqrt{t^2-|t^2-2t^{'2}|}$. 
 If $|t'|<|t|/\sqrt{2}$, this bound is $4\sqrt{2}t'$, which is proportional to the size of the time-reversal symmetry breaking term, as expected.  
In this model, a gap exists for any value of $t'\ne 0$, and this estimate proves the existence of this gap at least up to $|t'|=|t|$.
 
 \subsection{Green's function approach yields a gap}\label{Greens_func_yields_gap}
 To better understand the connection of this argument to interference and locality, consider the Green's function $G(j,i;\omega)=[(\omega-D)^{-1}]_{ji}$, the amplitude of the oscillations of the $j^\mathrm{th}$ site when the $i^\mathrm{th}$ is oscillated by a unit force at a frequency $\omega$. 
 This shows how far energy is transported at a given frequency. 
 We will show that this decays exponentially over a range of frequencies, implying that this range is a gap.~\footnote{Unlike the previous argument, here we cannot prove a vanishing density of states, but for disordered or amorphous systems, the mobility gap is the physically relevant feature, given that the density of states need not vanish in the gap.} 
 

First consider an ordinary insulator where there
 are two types of sites with different diagonal spring constants.  
Write $D=D_{diag}+\delta$, where $D_{diag}$ includes the diagonal pinning frequencies and $\delta$ includes only the off-diagonal ones.  
The Green's function can be expanded in a geometric series as $\frac{1}{\omega-D_{diag}}\sum_{n=0}^\infty \left(\delta   \frac{1}{\omega-D_{diag}}\right)^n$. 
 If $\omega$ is sufficiently far from the two diagonal entries, this series decays exponentially with distance. 
 Let us imagine ``bonds" drawn connecting any two sites with a non-zero value of $\delta$. 
 The first contribution to the sum for a pair of sites $i$ and $j$ appears at the $m^\mathrm{th}$ term if it requires $m$ steps to connect $i$ to $j$ along these bonds.  
 The magnitude of this term decays as  $(\max_{pq} |\delta_{pq}|/\min_{p} |\omega-D_{pp}|)^m$, with the rate of decay given by the ratio of the maximum off-diagonal component of $D$ to the minimum on-diagonal component of $\omega-D$.
 This suggests that the Green's function decays exponentially.
 This argument leaves out a factor that counts the number of paths connecting a given pair of sites, but when this factor is included, we find that the Green's function does in fact decay exponentially, as shown in Appendix~\ref{appendix:green_function}.
 Intuitively, with a short-range dynamical matrix, the network's energy transport is localized unless the couplings are large enough to be comparable with the difference between the applied frequency $\omega$ and the natural frequencies on the diagonal of $D$.
 
 For our system, this argument at first appears invalid since the off-diagonal couplings are not small when gyroscopes are strongly coupled. 
 A slight modification of this calculation does hold, however.
 For some forcing $\vec{f}$, the system responds with $\partial_t \psi_i=-i\sum_j D_{ij}\psi_j-i f_i$. 
 Through repeated differentiation, we can find the derivatives of $\psi_i$ in terms of the initial values of the $\psi$'s, and typically higher derivatives depend on the displacement of more distant gyroscopes. 
 This suggests that energy can be transported over long distances through chains of interacting sites.
 But if $D^2$ is almost diagonal, $\partial_t^2 \psi_i = -\sum_{jk} D_{ij}D_{jk}\psi_k-\sum_j D_{ij}f_j-i\partial_t {f_i}$, and the coefficient of $\psi$ on the right-hand side is $D^2=\omega_0^2\mathds{1}+\epsilon$, which is nearly diagonal. 
 Physically, the disturbance produced by a given gyroscope mostly reflects back to itself after two steps due to interference.
 The solution to the equation for forces at a frequency $\omega$ (using vector notation) is
 \begin{align}\label{eq:green_series}
 \psi&=\frac{1}{\omega^2-\omega_0^2-\epsilon}(\omega+D)f\nonumber\\
 &=\frac{\omega+D}{\omega^2-\omega_0^2}\sum_{n=0}^\infty \left(\frac{\epsilon}{\omega^2-\omega_0^2}\right)^n f,
 \end{align}
 which converges if $\epsilon$'s entries are small compared to $|\omega^2-\omega_0^2|$, so the Green's function (the coefficient of $f$) decays exponentially with the distance.


Though no part of the topological system is disconnected from the remainder by weak couplings, interference can nonetheless suppress tunneling of excitation from a given region.
This inteference can arise over short paths or long length paths, depending on the details of the network.
In the Hofstadter model we have used here, short paths of length two already generate significant interference,  enabling a simple test to identify the band gap.
In general, the interfering paths might be longer, requiring more and more terms in~\Eq{eq:green_series} to converge.
Though the paths may be long, the series will necessarily converge for a suitable choice of parameters (see Appendix~\ref{appendix:generalize_chern}).
If there is a gap, this method identifies its existence if~\Eq{eq:green_series} converges, even in the case of topological systems with strong off-diagonal couplings. 
This method also provides insights into the interference processes within the gap by following terms within the series.

We have discussed three cases. 
In trivial insulators with nearly-diagonal $D$, we can predict the presence of a gap from the diagonal elements. 
Systems dominated by bonding/anti-bonding behavior are instead nearly \textit{block} diagonal, and a gap will reside between bands of bonding and anti-bonding behavior.
Third, topological systems require strong coupling between sites to drive the system into a different phase -- one that cannot be described as a composite of non-interacting pieces.
In this strongly-coupled regime, interactions and eigenstates are non-local, threatening any prediction of band gaps without full diagonalization of the dynamical matrix~\cite{arovas_localization_1988,huo_current_1992}.
Nonetheless, we constructed a simple argument if $D^2$ is nearly diagonal. 
As shown in Appendix~\ref{appendix:generalize_chern}, this argument generalizes to a local method for finding locations of gaps even when $D^2$ is not nearly diagonal, since in such a case interference occurs between sufficiently long paths.

 Our argument applies both to topological systems and other systems with strong couplings for which $D^2$ is nearly diagonal in the bulk.
In topological phases, note that $D^2$ will \textit{not} be nearly diagonal on the boundary, and this delocalization signals the existence of protected edge modes.
For disordered and amorphous samples, the denominators in~\Eq{eq:green_series} could sometimes be small, but the total expression may nonetheless decay in a similar manner to the Green's function of disordered trivial insulators~\cite{anderson_absence_1958}.



\section{Real-space topology}\label{section:real-space_topology}

Now that we have gained insight into how gaps arise, we turn to characterizing their topology. 
In particular, we seek local, real-space probes for global topological invariants.
A natural starting point is the notion of flux pumping, which is a natural connection between a localized perturbation and a global consequence.
After observing how flux-pumping plays out in a mechanical context, we review the traditional notion of the Chern number and its real-space generalizations in the form of the Kitaev sum and the Bott index.
We compare these three methods of computing band topology and probe the extent to which we can approximate the Chern number using only the local structure of a network. 
Using a variant of the Kitaev sum, we find that, surprisingly, a purely local computation can approximate the real-space generalization of the Chern number.

\begin{figure*}
\includegraphics[width=\textwidth]{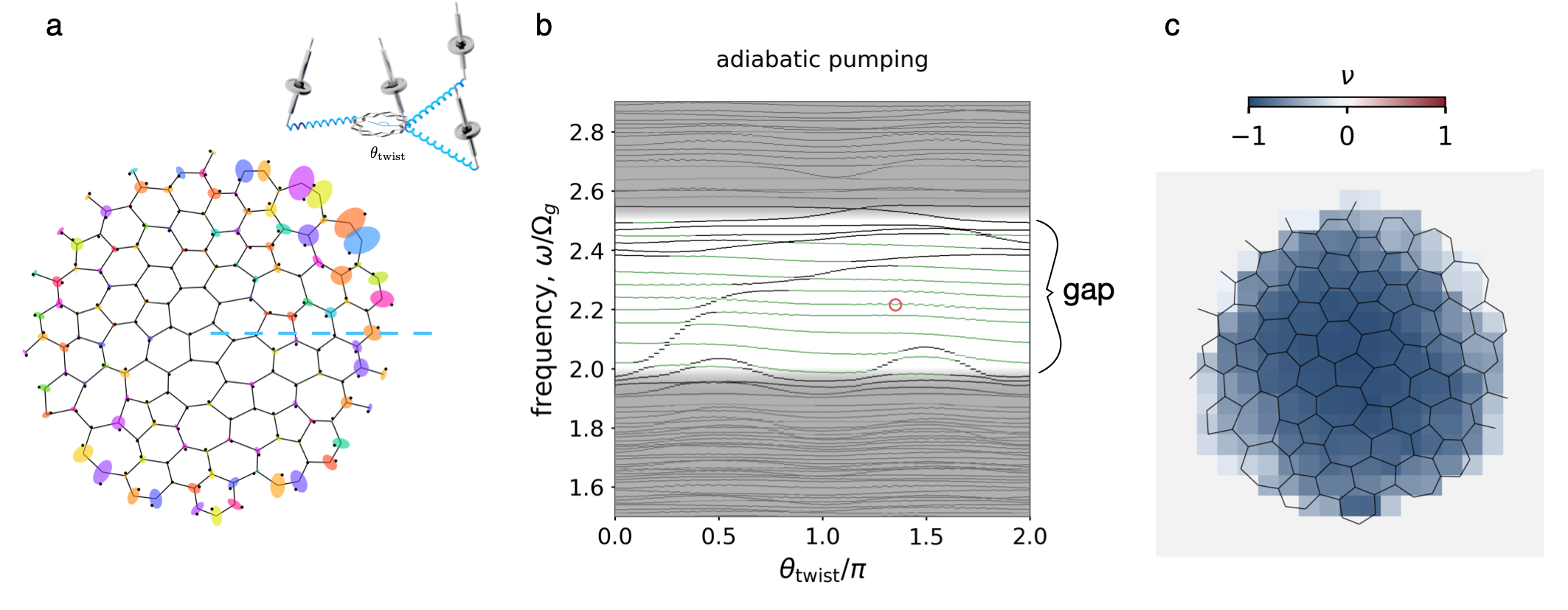}
\caption[]{
\textbf{The topological invariant in a small system can be measured by two related methods: the spectral response to twisted boundary conditions and the Kitaev sum.}
\textit{(a)} Twisting the bonds along the dashed blue line by attaching an extensible ring to each gyroscope above the dashed line drives a phenomenon akin to flux pumping from a threaded magnetic field.
The extensible ring, like a Hoberman ring, remains centered around the pivot point of the gyroscope and is free to rotate. 
This modulates the interaction by an angle, $\theta_{\textrm{twist}}$.
\textit{(b)} 
Upon twisting the interaction angle by $2\pi$ radians, the spectrum returns to its initial state, but the nontrivial topology ensures that one state from the central region is pumped from the lower band to the upper band, while states localized to the outer boundary (green curves) are pumped towards the lower band. 
Here, bands are shaded by a gray overlay. 
The red circle in the spectrum corresponds to the normal mode excitation shown in panel (a).
\textit{(c)} The Kitaev sum reveals that there is chiral conductance within the bulk of even a system $\sim 16$ gyroscopes in extent. 
Here, a region of summation half the size of the system is used to evaluate the sum at a grid of locations. 
The sum is proportional to the number of states at the system's core that are pumped from the conduction to the valence band. 
}
\label{fig_bulkboundary}
\end{figure*}


\subsection{Flux pumping as a local probe for band topology}

Arguably at the heart of band topology~\cite{laughlin_quantized_1981} lies the notion of flux pumping. 
For an electronic Chern insulator, threading magnetic flux through the core of an annulus effectively transfers occupancy of edge states from one edge to another. 
On one boundary, the process pumps electronic edge states into the bulk conduction band and bulk states in the valence band onto the edge, while on another boundary, the process pumps edge states into the valence band and pumps bulk states in the conduction band onto the edge.
With the goal of building a local description, we want to understand this topological phenomenon in terms of a localized perturbation.
A localized phenomenon that is similar to pumping flux through an annulus arises when threading one quantum of magnetic flux localized in the bulk of the material. 
This localized insertion gives rise to a quantized charge accumulation at the site of flux insertion~\cite{mitchell_amorphous_2018}. 

As magnetic flux is inserted into the bulk of the system, an induced electromotive force winds around the site. 
Given that a Chern insulator will exhibit chiral conductance, this induced field drives a current toward or away from the site of flux insertion, accumulating charge at the insertion site that is balanced at the outer boundary.
In the mechanical version of a Chern insulator, the states can similarly be transformed into each other via a mechanical version of flux pumping, wherein upper frequency states transform into lower frequency states via edge modes on the outer boundary for a topologically nontrivial network.
Since the magnetic flux is related to the phase winding of electrons, magnetic flux can be simulated by altering the interactions such that the force of one gyro on its neighbor is rotated by an angle $\theta_{\textrm{twist}}$:
\begin{equation}
F \sim \psi_i -\psi_j \rightarrow 
\psi_i - \psi_j e^{i\theta_{\textrm{twist}}}.
\end{equation}
To give a concrete picture of how this could be built in an experiment, we envision attaching an extensible Hoberman ring to a small number of gyroscopes, as illustrated in Figure~\ref{fig_bulkboundary}a.
Each ring remains centered about the hanging axis of its gyroscope, so that when a gyroscope is displaced, the ring uniformly extends in all directions.
The ring freely allows the gyroscope displacement to change in magnitude by expanding or contracting without any resistance and allows change in phase by rotating without any resistance.
To twist the gyroscope $i$'s response to its neighbor $j$, we affix the spring that couples $i$ and $j$ to a position on the ring offset from the gyroscope $i$'s tip. 
This introduces a phase-shifted torque on the gyroscope.
To mimic the effects of a gauge field, we are free to concentrate the modifications to the spring attachments along a radial cut of the annulus, shown as a blue dashed line in Figure~\ref{fig_bulkboundary}a,
effectively supplying a `twisted' boundary condition to the system along the cut of the annulus. 

Twisting the bonds along the cut deforms the resulting band structure, but the system returns to its original state once the twist has taken the value $\theta_{\textrm{twist}} = 2\pi$.
\Fig{fig_bulkboundary}b shows that during this operation, states localized to the outer boundary decrease in frequency, while a state localized to the center, where the effective magnetic flux resides, increases in frequency. 
The net result of the operation is the exchange of one state between the two bands. 
\Fig{fig_bulkboundary}b highlights this action in the gap: the edge (green) states are transformed into adjacent, lower frequency states, while the state at the center of the sample (purple) pumps from the lower band to the upper band.
In contrast, a trivial insulator has no exchange between the two bands, though extended states may be shuffled within each band separately.

Traditional treatment explains this phenomenon by looking at the band structure of the periodic bulk. 
We briefly review this more restrictive perspective before moving to a real-space point of view, where the lattice structure of the bulk is not needed and we are able to compute the band topology and understand flux pumping seen even for the small system shown in~\Fig{fig_bulkboundary}.

\subsection{Traditional view: Berry curvature}

The traditional point of view for computing band topology is to work entirely in reciprocal (momentum) space, which requires a crystalline material.
While more restrictive, we briefly lay out this formulation in order to introduce the projection operator and highlight the alterations required by symplectic symmetry.
Here, the Chern number for gyroscopes on a lattice is defined from the band structure of the equations of motion for small displacements.
Diagonalizing the matrix describing the network's dynamics yields $2N$ frequencies of the dispersion bands at each value of $\vec{k}$, where $N$ is the number of gyroscopes per unit cell.
For each value of $\vec{k}$, each band has a corresponding eigenvector, 
\eq |\boldsymbol{\psi}_j (\vec{k} )\rangle = \left(
\psi_1^R, ..., \psi_N^R,
\psi_1^L, ..., \psi_N^L
\right)
\qe
characterizing the amplitudes and phases of the $N$ gyroscopes' collective motion.
The symplectic symmetry of the dynamical matrix enables the eigenstates to be orthogonalized such that 
\begin{align}\label{symplectic_eigenstates_orthogonalized}
\langle \boldsymbol{\psi}_p |\cdot^\perp| \boldsymbol{\psi}_q \rangle &= 
\sum_\alpha
\xoverline{\psi^R_{p,\alpha}}  \, {\psi^R_{q, \alpha}}
- \xoverline{\psi^L_{p,\alpha}} \,  {\psi^L_{q, \alpha}}
 \\
&= \delta_{pq} \, \textrm{sign}(\omega_q), 
\end{align}
where $\alpha$ runs over each gyroscope and $\omega_q$ is the oscillation frequency of $| \boldsymbol{\psi}_q \rangle$.
In a lattice, momentum space is periodic, just as the real space lattice configuration is a tiling of the unit cell, implying that the Brillouin zone is a torus. 
Information about the connection of normal modes in each band is given by the Berry curvature, defined as
\eq 
\vec{\Omega}_n =  \nabla_{\mathbf{k}} \cross i \langle \boldsymbol{\psi} | \nabla_{\mathbf{k}} | \boldsymbol{\psi} \rangle.
\qe
This Berry curvature is akin to a magnetic vector potential in momentum space, affecting the motion of wavepackets as they traverse the system. 
The integral of the Berry curvature over the Brillouin zone describes the topological obstruction to a continuous connection between states in the band, encoded in the Chern number:
\eq 
\int_{BZ} \dif \mathbf{S} \cdot \boldsymbol{\Omega}(\mathbf{k}) =  2\pi \, C.
\qe 
The integer $C$ is the Chern number of the band~\cite{thouless_quantized_1982}.

\begin{figure}
    \centering
    \includegraphics[width=\columnwidth]{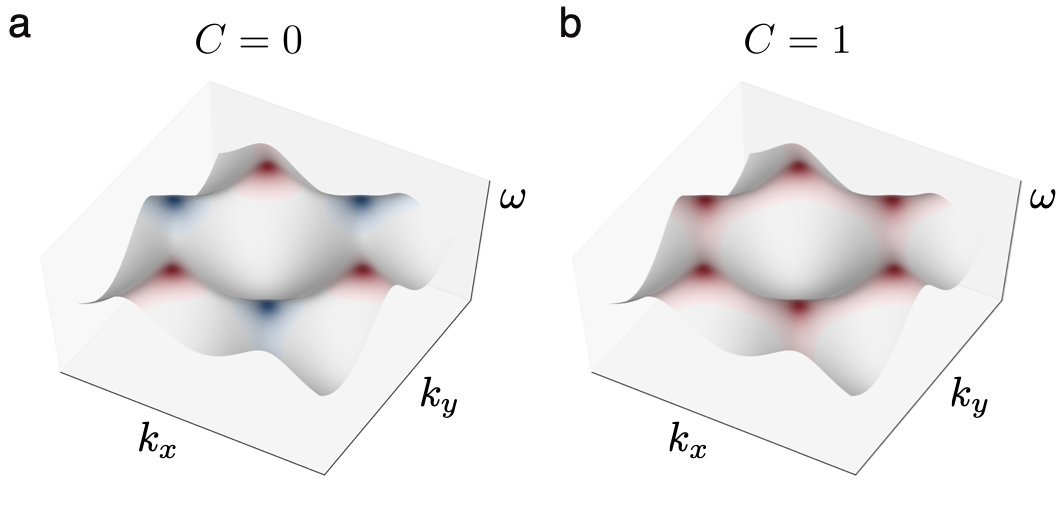}
    \caption{
    \textbf{The Chern number is given by the integreated Berry curvature, which describes the connection between states in a band.}
    \textit{(a)} A topologically trivial band may have nonzero Berry curvature, but its integrated Berry curvature vanishes.
    \textit{(b)} For a band with Chern number of $C=1$, the integrated Berry curvature sums to $2\pi$.
    }
    \label{fig:intro_berry_curvature}
\end{figure}

\begin{figure}
    \centering
    \includegraphics[width=\columnwidth]{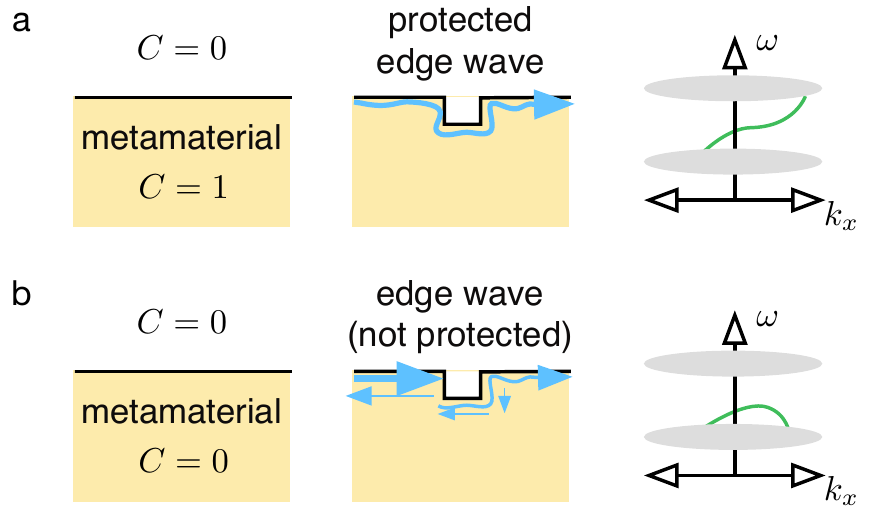}
    \caption{\textbf{The Chern number counts the number of modes occupying a band gap.}
    \textit{(a)} In a system with a single gap and a single boundary, the number of right-moving modes --- ie, sets of possible eigenstates with positive group velocity --- minus the number of left-moving modes is determined by the Chern number of a band. 
    Nonzero Chern numbers give rise to topologically protected chiral edge modes, which are robust against back-scattering.
    \textit{(b)} A topologically trivial insulator does not support protected edge modes on the boundary.
    }
    \label{fig:intro_chern_number}
\end{figure}

\Fig{fig:intro_berry_curvature} shows two schematic examples of a band colored by its Berry curvature. 
In~\Fig{fig:intro_berry_curvature}a, the Berry curvature is nonzero, but its integral over the Brillouin zone vanishes: this is a trivial insulator with inversion symmetry breaking.
In~\Fig{fig:intro_berry_curvature}b, however, the contributions add, and the total value is a nonzero multiple of $2\pi$.
This nonzero Chern number signals the existence of the topologically protected chiral modes that live on the boundary of a system~\cite{thouless_quantized_1982}.
Topological protection ensures that, in the absence of available counter-propagating states, an edge wave will not scatter, passing around inclusions or voids and readily changing direction along jagged boundaries (\Fig{fig:intro_chern_number}).

An alternative representation of the Chern number uses the phase-invariant formula~\cite{avron_homotopy_1983}
\begin{equation}
\label{chern_num}
C_{j} = \frac{i}{2 \pi} \int \textrm{d}^2 k\ \mathrm{Tr}[\partial_{k_\alpha} P_j  P_j \partial_{k_{\beta}} P_j \varepsilon^{\alpha \beta}],
\end{equation}
where $C_j$ is the Chern number of the $j^{\textrm{th}}$ band, $\varepsilon^{\alpha\beta}$ is the antisymmetric Levi-Civita symbol, and $P_j$ is the projection matrix defined for our system as
\eq \label{projector_symplectic}
P_j = | \psi_\alpha \rangle 
\langle \psi_\alpha | 
Q \, \textrm{sign}(\omega_j),
\qe 
where
\eq  
Q = \left( \begin{array}{cc}
\mathbb{I}_{n}  & 0 \\
0  &  - \mathbb{I}_{n}
\end{array} \right) .
\qe
The factors of $Q \, \textrm{sign}(\omega_j)$ arise from the symplectic structure of gyroscopes' equations of motion and the normalization of the states in~\Eq{symplectic_eigenstates_orthogonalized}.
If the symplectic symmetry were ignored, and the projector was instead defined using a simple outer product of bands which were not orthonormal, a non-physical distribution of Berry curvature would result, though this does not affect any of the computed Chern numbers.
By using the symplectic formulation, the correct Berry curvature distributions are readily obtained.

\begin{figure*}
\includegraphics[width=\textwidth]{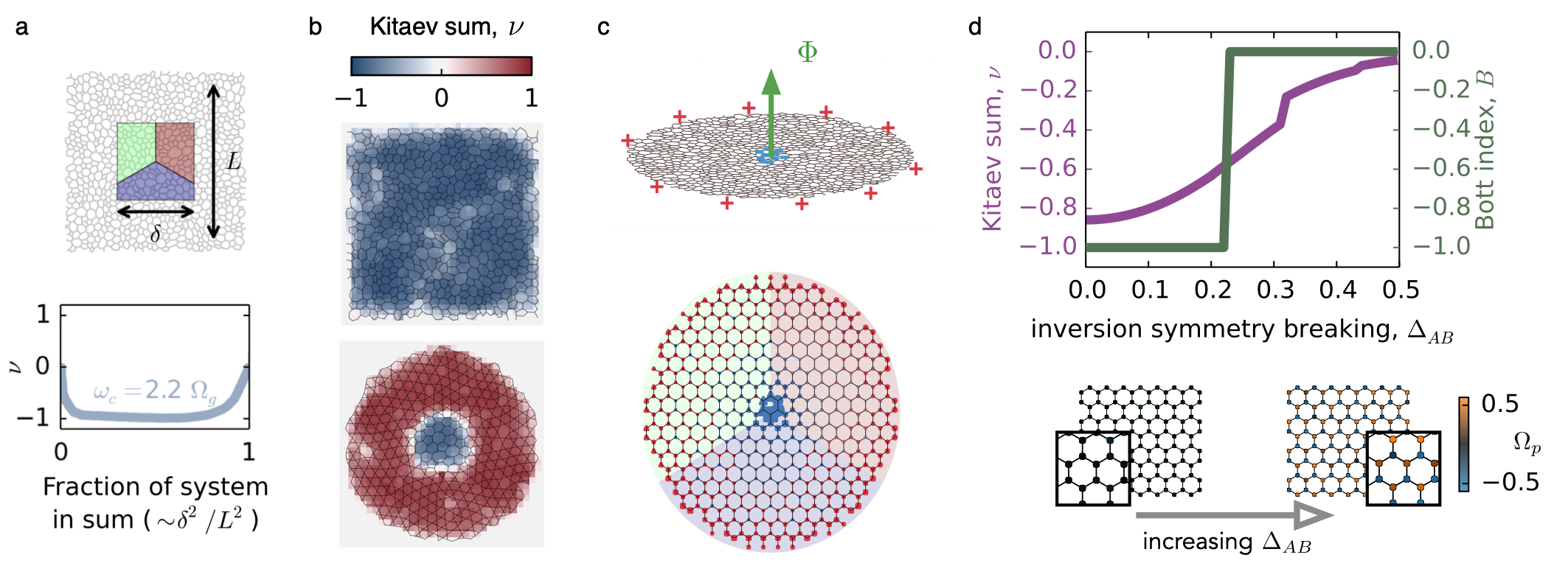}
\caption[]{
    \textbf{The Kitaev sum and Bott index both measure band topology in real space.}
\textit{(a)} The Kitaev sum measures the band-integrated spectral flow at a particular location in the bulk of the material.
The sum converges to a target integer value corresponding to the Chern number of the system in the limit of an infinitely large system and large region of summation that does not enclose a boundary. 
The blue curve shows the convergence of $\nu$ for increasing sizes of the summation region $\delta$. 
The projector used for this calculation maps all states of the amorphous system that are above $\omega_c=2.2\Omega_g$ to themselves and maps states below $\omega_c$ to a zero vector. 
\textit{(b)} Centering the summation region on different locations in an amorphous network in which all sites are trivalent (top) or in which two different local geometries are present (bottom) generates a spatially-resolved measurement of the local topological character of the material.
\textit{(c)} In electronic systems, the Kitaev sum measures the net charge that accumulates near the location of magnetic flux insertion (top panel)~\cite{mitchell_amorphous_2018}. 
In gyroscopic and electronic systems alike, the contribution to the sum is dominated by sites close to the site of flux insertion, represented by the large blue markers at the center of the circular sample. 
The size of each site's marker is proportional to the magnitude of the sum of terms in the Kitaev sum that include that site. 
Blue markers indicate a negative net contribution, while red terms indicate a positive contribution. 
An equal and opposite contribution resides on the boundary sites, mimicking the opposite charge that resides at the boundary in the electronic case, represented by the red markers at the edge of the circular sample. 
\textit{(d)} The Bott index measures a topological obstruction preventing band-projected position operators from commuting. The index is a global measurement of band topology and requires closed boundary conditions.
We contrast the Kitaev sum with the Bott index by studying a small gyroscopic honeycomb network subjected to inversion symmetry breaking. 
The Kitaev sum here has summation window set to include 45\% of the sites in the system.
The inset below in panel \textit{(d)} shows the pinning strength at each site (ranging from blue to orange) for the samples used in the calculation with open (periodic) boundary conditions for the Kitaev sum (Bott index). 
The pinning strengths $\Omega_p$ of each site are disordered by adding an amount $V$ sampled from a Gaussian distribution with a standard deviation of $0.1\,  \langle{\Omega_p}\rangle$, and the gyroscopes have pinning strengths biased by their sublattice site: $\Omega_p = \langle \Omega_p \rangle(1 \pm \Delta_{AB}) + V$. 
}
\label{fig_kitaev}
\end{figure*}

\subsection{Real-space generalizations of Chern number}
If disorder is present in a lattice or if a gyroscopic network has amorphous structure, translational symmetry is broken, and the Chern number previously described cannot be defined due to the absence of a Brillouin zone. 
However, several real-space generalizations to the Chern number have been proposed. 
We discuss two approaches here.
We use both to investigate topological phase transitions, and one will even enable us to build an approximation to the Chern number using only the local character of the network.

\subsubsection{Kitaev sum}

The Kitaev sum generalizes the notion of Chern number to a real-space measure of chirality integrated over a band in the bulk of a 2D material~\cite{kitaev_anyons_2006,mitchell_amorphous_2018,ma_hamiltonian_2017}.
By defining a projection operator, $P$, mapping states above a cutoff frequency, $\omega_c$, to themselves and states below the cutoff to zero, we obtain a real-valued number $\nu$ via
\begin{equation}
    \nu(P) = 12 \pi i \sum_{j \in A} \sum_{k \in B} \sum_{l \in C} 
            \left(
            P_{jk} P_{kl} P_{lj} - P_{jl} P_{lk} P_{kj}
            \right).
\end{equation}
The projection operator could instead be defined to map states \textit{below} the cutoff $\omega_c$ to themselves, as is often done in electronic systems; in this case, $\nu \rightarrow -\nu$.
When negative frequency bands are included, it is important to include the symplectic factors in~\Eq{projector_symplectic}.
$A$, $B$, and $C$ are three non-overlapping summation regions defined in a counterclockwise fashion as shown by the red, green, and blue regions of~\Fig{fig_kitaev}a and c, and these can be of any shape. 

In electronic systems, the Kitaev sum measures the charge that accumulates for a localized magnetic field flux inserted into the bulk position where the three regions $A$, $B$, and $C$ meet.
For a single magnetic flux quantum, the amount of charge is $\nu e$.
In our gyroscopic system, $\nu$ similarly relates directly to flux pumping, signifying the number of states exchanged between the boundary and the site of flux insertion shown in~\Fig{fig_bulkboundary}a~\cite{mitchell_amorphous_2018}.

One feature of the Kitaev sum is that the result is not strictly integer-valued.
In the limit that the region of summation encloses many gyroscopes without enclosing the material's boundary, however, the value of $\nu$ does converge towards an integer.
This non-integer character of $\nu$ is useful in that the rate of convergence indicates the localization length of modes at the cutoff frequency: $\nu$ approximates its target value once the region of summation is larger than the localization length.


\subsubsection{Bott Index}

An alternative approach which rests on $K-$theoretic methods is to measure what has become known as the Bott index~\cite{loring_disordered_2010}.
Like the Chern number, a nonzero Bott index signals the impossibility of finding a complete, orthonormal basis of localized functions (`Wannier states') among the states included in the projection operator.
Unlike the Kitaev index, the Bott index requires the construction of a periodic system, so translational invariance is artificially restored on some scale, though this scale can be made arbitrarily large. 
Given a fully periodic sample of spatial extent $W$ in both spatial dimensions $X$ and $Y$, the positions of each gyroscope are indexed by $\Theta = 2\pi X / W$ and $\Phi = 2\pi Y / W$.
Equipped with a projection operator $P$ as before, we define band-projected position matrices 
\begin{equation}
    P e^{i \Theta}P \sim \left(
        \begin{array}{cc}
        0 & 0  \\
        0 & U
        \end{array}
    \right)
    \hspace{30pt}
    Pe^{i \Phi} P \sim \left(
        \begin{array}{cc}
        0 & 0 \\
        0 & V
        \end{array}
    \right),
\end{equation}
with the right hand sides expressed in the basis of eigenmodes.
If the cutoff frequency of $P$ lies in a mobility gap, the block nonzero components $U$ and $V$ are almost unitary and will almost commute~\cite{loring_disordered_2010}. 
The Bott index, $B$, defined as 
\begin{equation}\label{eqn_bott_index}
    B = \frac{\Im \left[ \textrm{Tr} (\log(VUV^\dagger U^\dagger)) \right]}{2\pi i},
\end{equation}
indicates if $U$ and $V$ are `close' to a different pair of matrices which are exactly unitary and exactly commute.
Since the logarithm is multivalued,~\Eq{eqn_bott_index} must be defined more precisely: we evaluate the sum of the logarithm of the eigvenvalues of $VUV^\dagger U^\dagger$ and take the imaginary components to be as small as possible. 
A potential advantage of the Bott index over the Kitaev sum is that it returns an exact integer, rather than converging towards an integer value, at the cost of requiring the introduction of (potentially artificial) periodicity to the sample. 
A potential disadvantage is that the Bott index measurement is global: a given sample returns a single value, whereas the Kitaev sum can be performed in a spatially-resolved manner (\Fig{fig_bulkboundary}b and~\Fig{fig_kitaev}a-c).
For a system composed of multiple patches of different topological phases, such as in the lower panel of~\Fig{fig_kitaev}b, the Bott index selects only a single global readout.


\subsection{Determining the Chern number by Local Approximations}
Given that such small systems can still be used to accurately compute real-space topological invariants and exhibit bulk-boundary correspondence, it is tempting to think that we can use the local properties of a network to approximate the topological invariant, under the assumption that the region is not exceptional.
Here, we will build on Kitaev's insight to address this challenge for systems that satisfy the same conditions as we required in section~\ref{section:understanding_gap_in_regime}: namely, that (1) some off-diagonal elements of $D$ are large compared to on-diagonal elements, and (2) the square of the dynamical matrix is nearly diagonal such that $D^2 \approx \omega_0^2 \mathds{1} + \epsilon$. 
The goal is to approximate the projection operator, which is short-ranged~\cite{kohn_theory_1964,hastings_lieb-schultz-mattis_2004}.
This is a critical assumption behind the classification of interacting topological phases, which -- like amorphous materials -- are studied in real-space rather than momentum space. 

To begin, we use an expression we derived in~\cite{mitchell_amorphous_2018} that casts Kitaev's formula into
\begin{equation}\label{eq:kitaev_mod}
\nu =\mathrm{Im\ } 8\pi\sum_{ijk} P_{ij}P_{jk}P_{ki},
\end{equation} 
where $P$ is the projection onto states below a given frequency, and the sum is over all sets of three points in the network that form a triangle surrounding a given point $q$ in a counterclockwise orientation. 
This formula can be derived as the charge that accumulates when a magnetic flux quantum is threaded through $q$, and this expression translates directly into Kitaev's (and vice versa) if $P$ is the exact projector. 
However, we will use an approximation to the projector, and~\Eq{eq:kitaev_mod} seems more reliable in the situation when $P$ is known only approximately. 
Kitaev's formula requires a choice of subregions $A$,$B$,$C$ that can change the value of the sum if $P$ is merely approximated, but our expression requires no such choice.

\begin{figure}
\includegraphics[width=\columnwidth]{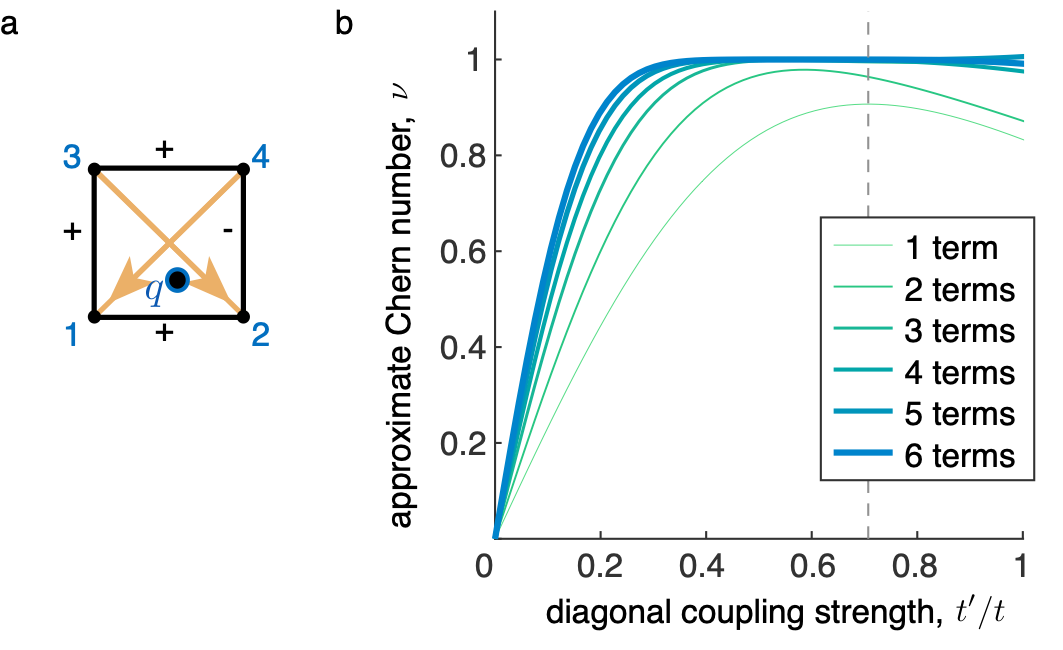}
\caption{
\textbf{The Chern number can be approximated by a purely local, real-space calculation.}
\textit{(a)} In the Hofstadter model, there are two three-hopping triangles formed from nonzero entries in $D$ that encircle $q$, which are needed to calculate the lowest-order approximation to the Chern number. 
The two triangles are 123 and 124.
\textit{(b)} The approximation to the Chern number (\Eq{eq:kitaev_mod} and \ref{eq:projector_series_expansion}) converges toward the true value of $\nu=1$ as the number of terms included from the expansion of~\Eq{eq:projector_series_expansion} grows.
The thin green curve corresponds to the first term, given by~\Eq{eq:approx_chern}, and thicker lines varying from green to blue indicate the result from including two, three, four, five, or six terms from~\Eq{eq:projector_series_expansion}. 
The measurement is shown here as a function of the complex, next-nearest-neighbor hopping strength $t'/t$ (yellow arrows in panel a). 
The dashed, vertical gray line indicates the value of the ratio $t'/t = {1}/\sqrt{2}$, where the approximation should be the most accurate. 
Above $t'>t$, the series does not converge for our choice of $\omega_0$ in $D^2 = \omega_0^2\mathds{1} + \epsilon$.  
Note that the single-term approximation (thin green curve) reaches its maximum at this coupling strength, and further terms improve the result near $t'/t = {1}/\sqrt{2}$.
\label{fig:approximate_chern}}
\end{figure}

The projection operator (or `projector') in a coordinate system where $D$ is diagonal is $\sum_{\omega_n<\omega} |n\rangle\langle n|$, where $\omega_n$ are the frequency eigenvalues and $|n\rangle$ the eigenvectors.  
This is an inherently nonlocal object, but we can overcome this difficulty by writing the sum in an algebraic form as $\sum_{\omega_n} \frac{1}{2}(1-  \omega_n/\sqrt{\omega_n^2})|n\rangle\langle n|$, where the square root is the positive square root.  
Recast in a coordinate-independent way,  $P = \frac{1}{2}(\mathds{1}- D / \sqrt{D^2} )$ which can be expanded using the assumption $D^2=\omega_0^2 \mathds{1}  +\epsilon$, where $\epsilon$ is small, as
\begin{equation}\label{eq:projector_series_expansion}
P=\frac{1}{2}\mathds{1}-\frac{D}{2\omega_0}\sum_{n=0}^\infty\left(-\frac{\epsilon}{4\omega_0^2}\right)^n \binom{2n}{n}.
\end{equation}
Here, we have used the Taylor series $\frac{1}{\sqrt{1-x}}=\sum_{n=0}^\infty \binom{2n}{n} (x/4)^n$.
Substituting into~\Eq{eq:kitaev_mod}, we note first that the first term ($\frac{1}{2}\mathds{1}$) and its cross-terms with the other factors contribute only real terms to the
sum because $D$ is Hermitian and can therefore be neglected. 
Only the sum remains.
If the $n=0$ term dominates, then
\begin{equation}\label{eq:local_chern}
\nu \approx\mathrm{Im} \frac{\pi}{\omega_0^3}\sum_{ijk} D_{ij}D_{jk}D_{ki}.
\end{equation}
This is a local formula for the Chern number. 
Since $D_{ij}$ is short-ranged, the number of triangles $ijk$ enclosing a point that contribute nonzero terms is limited.

A simple model both provides intuition and tests the accuracy of this result.
For the Hofstadter model we considered before, which is associated with a gyroscopic network in the weakly-interacting limit $\Omega_k/ \Omega_p \ll 1$, $\nu$ is proportional to the product of the magnitudes of the couplings around the sides of a triangle times the sine of the net phase factor, normalized by $\omega_0^3$.
Intuitively, $\omega_0$ is akin to a mean value of the coupling magnitudes, so this cancels off much of the dependence on the overall magnitude of the couplings, leaving mainly the dependence on the phases.
For the square model, take $q$ to be the point shown in~\Fig{fig:approximate_chern}a.  
There are only two triangles surrounding $q$, which each contribute $it^2 t'$.  
They are each counted three times, since the three vertices $ijk$ can be enumerated with any cyclic permutation while retaining the proper counter-clockwise ordering, so 
\begin{equation}\label{eq:approx_chern}
\nu \approx \frac{3\pi t^2 t'}{4(t^2+t^{'2})^{\frac32}}.
\end{equation}
A true topological invariant should be quantized, but clearly this expression is not. 
We expect the approximation to work best when $t'$ is close to $t/\sqrt{2}$ so that the next-nearest-neighbor entries in $D^2$ cancel.  
At this point, the approximation is $\pi/ (2\sqrt{3})\approx .9$, which is close to 1.
A plot of the approximate Chern number against a range of complex hopping magnitudes $t'$ in~\Fig{fig:approximate_chern} is nearly constant for a range of values of $t'$ near $t/\sqrt{2}$.
Including more terms in the sum in~\Eq{eq:local_chern} increases the accuracy of the Chern measurement in the vecinity of $t'=t/\sqrt{2}$. 
As discussed in  Appendix~\ref{appendix:generalize_chern}, convergence of the series requires $\omega_0$ to be chosen so that $\omega_0^2$ is at least as large as half the maximum eigenvalue of $D^2$.
The method discussed here for computing a local approximation to $\nu$ generalizes to any system with a gap, regardless of whether $D^2$ is nearly diagonal, as we show in Appendix~\ref{appendix:generalize_chern}.

This approximation to the real-space Chern invariant does not require a full diagonalization of the dynamical matrix. 
This result expands our computational machinery not only to predict if a gap will appear (as derived before), but also to predict if that gap will be endowed with chiral edge modes.

\begin{figure}
\includegraphics[width=\columnwidth]{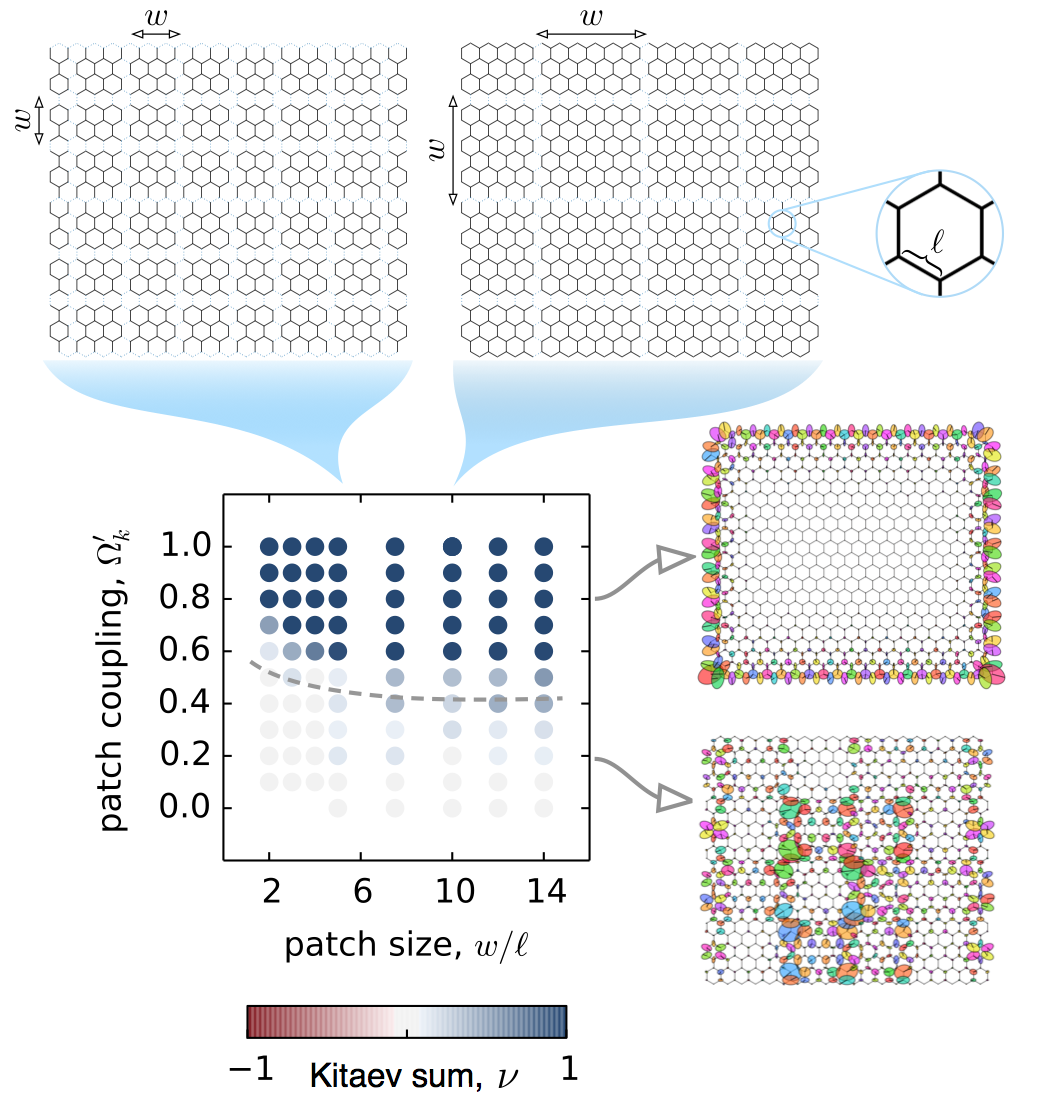}
\caption[]{
\textbf{In order for patches of Chern insulator to constitute a topological sample, the patches must be sufficiently coupled}.
By varying the spring constants connecting initially uncoupled regions of a gyroscopic network, we find that a threshold coupling strength is required to create a macroscopically topological sample.
Here, we used a $21\times21$-cell gyroscopic honeycomb configuration with $\Omega_g = \Omega_k$ and open boundary conditions. 
On the right side, two example eigenmodes with $\Omega'_k =0.2\Omega_k$ and $0.8 \Omega_k$ show the change in spatial structure of the excitations nearest $\omega = 2.25 \Omega_p$ for patch sizes $w/\ell=7.5$, where $\ell$ is the bond length. 
Below a critical patch coupling strength, edge modes span the boundaries between patches throughout the material, whereas above a critical patch coupling strength, a single edge mode resides on the outer rim of the network.  
The localization length $\xi$ for a mode in the middle of the band gap is on the order of the bond length: $\xi\approx \ell$, with the precise value depending on the orientation of the boundary with respect to the lattice.
So long as the patch size is greater than $\sim 2 \xi$, the transition from trivial to topological occurs at a fixed coupling strength.
 }
\label{fig_homogeneity}
\end{figure}

\subsection{Mesoscopy vs homogeneity in determining Chern number}\label{subsection:mesoscopy_vs_homogeneity}

We have found that even in relatively small systems, a topological invariant can be readily approximated by the Kitaev sum without periodic boundary conditions.
This demonstrates that band topology can be encoded on a mesoscopic scale -- far smaller of a scale than has been traditionally investigated.
If multiple topological patches are adjacent but uncoupled, then each patch will have its own edge states, but the system as a whole will not behave as a Chern insulator, since these states are distributed throughout the bulk. 
Since the `edge modes' of each patch occupy a frequency range that corresponds to the gap of a fully connected system, the material as a whole does not behave as an insulator. 
What degree of coupling between patches is required for these patches to behave as a single insulating system with chiral modes only at its outer boundary, rather than throughout its interior?
Here we study the transition of weakly coupled mesoscopic patches of Chern insulator into a single unit of material.

\Fig{fig_homogeneity} shows patches of gyroscopic honeycomb networks bonded together by spring coupling with variable strength (dashed bonds in~\Fig{fig_homogeneity}). 
With these bonds completely floppy ($\Omega_k' \ll \Omega_k$), the composite system is a trivial insulator, with a real-space Chern number measurement (here using the Kitaev sum) of $\nu \rightarrow 0$ (gray circles in the phase diagram of~\Fig{fig_homogeneity}).  
If the coupling is strengthened, however, the material transitions to a Chern insulator at $\Omega_k' \approx 0.4 \Omega_k$.
Notably, so long as the patches are sufficiently large that the edge modes localized to their boundaries do not span the entire patch, the value of $\Omega_k'$ at the transition does not vary with patch size.
\Fig{fig_homogeneity} shows the associated phase diagram: for cell sizes larger than $w \gtrsim 2 \ell$, the transition from trivial ($\nu = 0$) to topological ($\nu = -1$) occurs at a constant coupling strength.
This is an extension of our earlier observations in~\Fig{fig:spectrumflow} in which varying the coupling between unit cells led to a topological phase transition at a critical coupling strength.
There, a sufficiently strong coupling ($\Omega_k' \approx 0.4 \Omega_k$) was necessary to transform a trivial insulator of bonding/anti-bonding states into the delocalized states of a Chern insulator. 
Here, in addition, we find that even if large patches individually register as topological, a significant coupling is still required for the sample as a whole to have a nonzero topological invariant. 

Below a critical coupling, the conductance of this gyroscopic material is not insulating on account of interior interfaces between insulating grains.
This highlights the need for homogeneity in the coupling for a topological material to insulate throughout its bulk
and highlights one of the ways in which topological insulation can break down in the presence of heterogeneity.

In the process of establishing a local origin of topological band gaps, we encountered the importance of homogeneity in the strengths of bonds throughout the bulk.
Violating this assumption of homogeneity in couplings can change a material's topological phase.
We now turn to the other way of breaking homogeneity in the bulk: modulating pinning strengths at each site. 
Consistent with our findings so far, we find that whereas inhomogeneity can break topological insulation, spatial order is utterly irrelevant to topological order.
Indeed, in the final section of this article, we find that even in their response to on-site disorder, amorphous structures behave like their crystalline counterparts and undergo localization transitions consistent with the same scalings as in ordered crystals.




\section{Topology and Anderson localization in gyroscopic metamaterials}


Topological insulators are famous for the chiral waves living at a system's edge at frequencies in the gap.
However, there are also differences in the bulk modes at frequencies in bands bounding the gap.
These modes can conduct energy in phononic metamaterials, where there is no Pauli exclusion and all modes are available excitations.
For topologically trivial materials in 2D, any disorder is sufficient to localize all modes in a large enough system. 
For Chern insulators, on the other hand, some normal modes in bands will be spatially extended across the entire system~\cite{huo_current_1992}.
The study of how states localize as disorder increases or as the system size grows -- so-called `Anderson localization' -- has provided a powerful framework to understand and predict the conductivity of a wide range of materials~\cite{anderson_absence_1958,abrahams_scaling_1979,Lagendijk_Fifty_2009}.
Recent extensions incorporate topological order into this scheme~\cite{xue_quantum_2013}.


How do \textit{amorphous} Chern insulators behave when subjected to strong disorder? 
In the first half of this section, we study the scaling of the topological index with system size, with particular attention to the amorphous case.
By varying the disorder, we find  behavior consistent with the expected scaling for symmetry class A, including annihilation of extended states in the mobility gap at the transition. 

We then use inversion symmetry breaking to add a second dimension to this phase transition and find an avenue for disorder to drive a trivial insulating phase into a topological one.
This re-entrant topological behavior, dubbed `topological Anderson insulation', is surprising, since typically disorder drives systems away from the topological phase~\cite{groth_theory_2009,stutzer_photonic_2018,meier_observation_2018}.
Using experiments and simulations, we identify this interesting feature of the interplay between disorder and topology in the mechanical context for the first time.

\begin{figure}
    \includegraphics[]{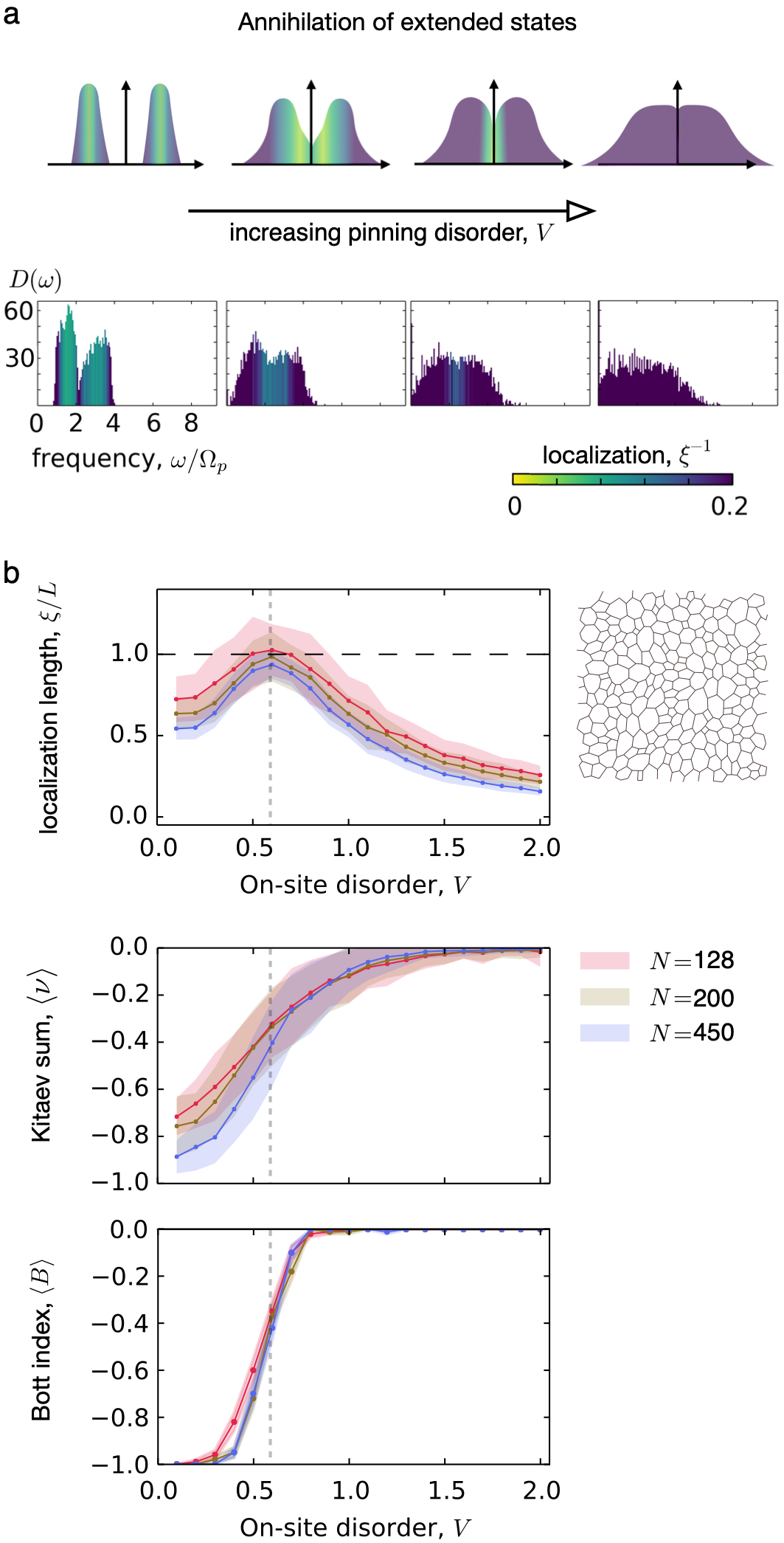}
    \caption[]{
    \textbf{Increasing disorder strength drives extended states towards the gap, where they annihilate as the system passes from topological to a trivial insulator.
    }
    \textit{(a)} Illustration of the annihilation of extended states (yellow states) in the gap as pinning disorder strength $V$ increases, as shown in a schematic cartoon (upper panel) and in computed spectra for a collection of 20 disordered amorphous gyroscopic networks of 450 gyroscopes (lower panel) with $\Omega_k = \langle\Omega_p \rangle$.
    The colors of these spectra reflect the localization of states, $\xi$, measured in units of inverse average bond length. 
    These extended states (yellow) carry the Hall conductance for the system, so their annihilation triggers a transition to a trivial insulating phase, in which all modes are localized (purple).
    \textit{(b)} As disorder grows, the localization length in the gap grows to the system size, aligning with the change in average topological invariant of the top band from -1 to 0. The vertical dashed line marks the disorder strength at which the topological transition occurs.
     }
    \label{fig_levannihilation}
\end{figure}


\subsection{Anderson Insulator transition}
Strong disorder in pinning strengths drive a transition to the trivial insulating regime.
In the canonical Anderson localization picture without any topological effects, the addition of infinitesimal disorder localizes all modes in a 2D system. 
In the presence of nontrivial topology, however, stronger disorder strengths are required to drive Chern insulators into the trivial insulating state. 
For small disorder, each band acquires tails of localized states, and bands carrying Hall conductance are confined to a region near the center of each band (leftmost panels in~\Fig{fig_levannihilation}a).
As the disorder strength grows, these states approach and annihilate in a frequency region which was previously a mobility gap. 
This process is called `annihilation and levitation' due to the annihilation of the extended states in the gap and the subsequent rise in variance in the spacing between eigenfrequencies.
In \Fig{fig_levannihilation}, we show that amorphous gyroscopic networks display this behavior as well. 

\Fig{fig_levannihilation}a shows this annihilation of extended states for 
amorphous networks of 450 gyroscopes.
Without pinning disorder, the two bands of states which span the system are separated by a mobility gap. 
As we increase random pinning frequencies $\Omega_p$ drawn from a Gaussian distribution with width $V$, however, the extended states move toward the gap, meet in the frequency range which was previously the middle of the gap, and annihilate, shown in~\Fig{fig_levannihilation}a.
\Fig{fig_levannihilation}b shows that the localization length in the middle of the gap ($\omega \approx \Omega_p + 1.25 \Omega_k$) rises as the extended states invade.
The localization length peaks at a value near the system size, then falls as disorder dominates.
The peak in localization length coincides precisely with the change in the ensemble-averaged topological index -- whether Kitaev sum or Bott index -- from nonzero to zero. 
Thus, the annihilation of the extended states mark the transition from a topological to trivial phase.

\begin{figure*}[ht]
\includegraphics[width=\textwidth]{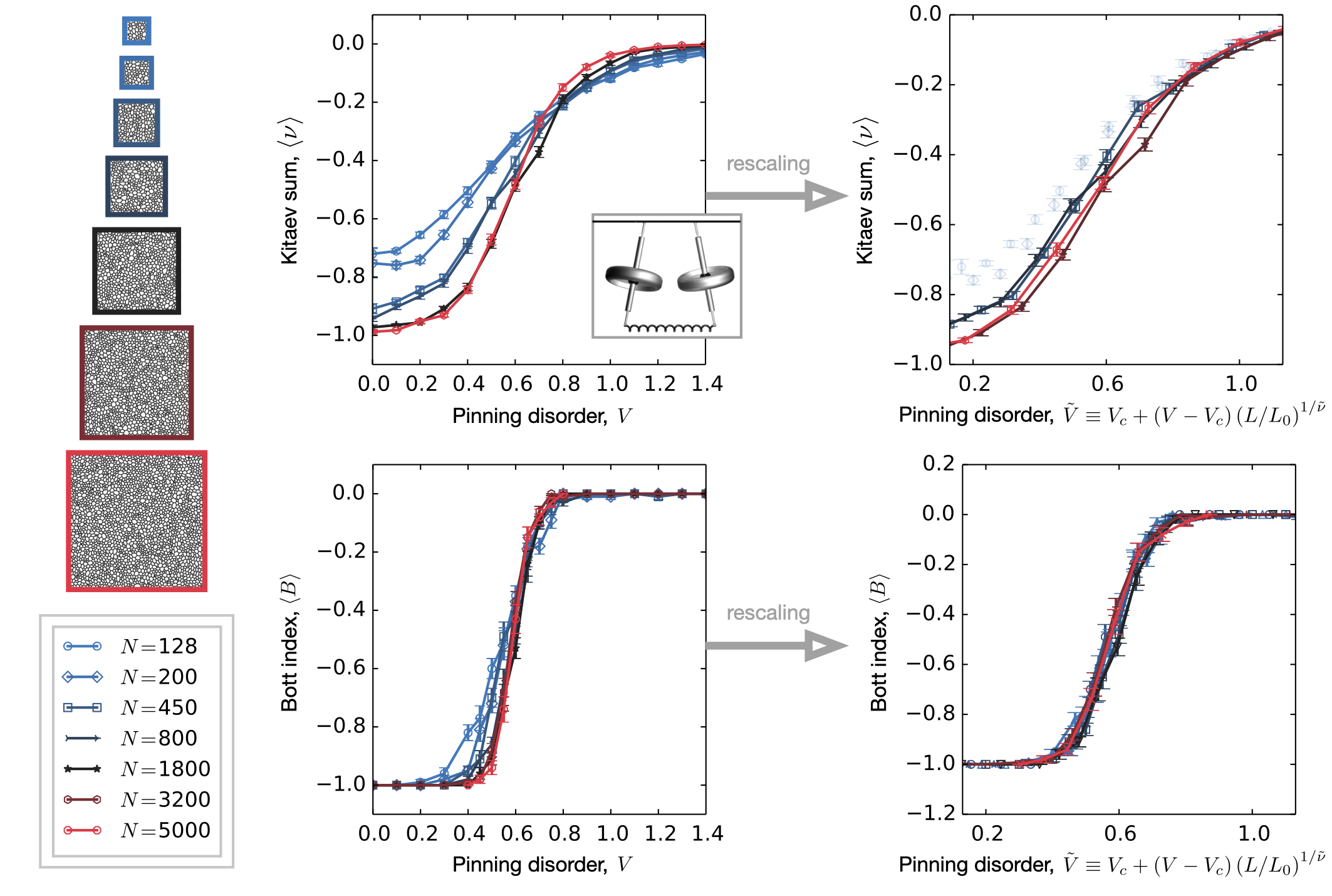}
\caption[]{
\textbf{Scaling collapse of the Anderson localization transition in our gyroscopic Chern insulators is consistent with the universality class of Cartan label A.
}
As disorder is increased, the topological index of the upper band averaged over many samples smoothly varies from -1 to 0. 
The spatial structures of networks used here (left column) are Voronoi tessellations of hyperuniform point sets generated as in Ref.~\cite{mitchell_amorphous_2018}, with a gyroscope placed at every vertex and bonds lining the polygons of the tessellation. 
As the size of the system for which the index is computed grows (blue to black to red in each panel), the transition grows sharper. 
Each data point represents the average Kitaev sum $\langle \nu \rangle$ or Bott index $\langle B \rangle$ for 100 amorphous gyroscopic networks (ten disorder realizations for each of ten randomly-generated amorphous structures) subjected to on-site disorder strengths modulating $\Omega_p$ at each site by an amount taken from a Gaussian distribution.
The pinning disorder strength is measured in units of $\langle\Omega_p\rangle$, and we let $\Omega_k=\langle\Omega_p\rangle$.
Rescaling the disorder values according to a power law of the system size collapses the curves with an exponent $\tilde{\nu} = 2.6$ and $V_c = 0.63 \, \langle \Omega_p \rangle$, as shown in the right panels. 
}
\label{fig_scaling}
\end{figure*}

By measuring the topological invariant for systems of different sizes, we find sharper and sharper transitions to the trivial phase as the system size increases. 
Intuitively, a disordered topological system whose size is comparable to the localization length will be susceptible to random variations in disorder such that different realizations of disorder may yield a trivial phase. 
Near the transition, where the localization length of modes in the center of the gap is large, proportionally larger systems are required to suppress fluctuations in the resulting topological invariant. 
This is the case both for amorphous structures shown in~\Fig{fig_levannihilation} and~\ref{fig_scaling}, as well as for the honeycomb lattice we studied earlier.
If the transition is perfectly sharp in the limit of an infinitely large system, then the system with disorder strength $V < V_c$ is a Chern insulator supporting topologically protected chiral edge modes, while systems with $V > V_c$ will register $\langle \nu \rangle \rightarrow 0$ and $\langle B \rangle \rightarrow 0$.

In~\Fig{fig_scaling}, we compute the topological invariants for thousands of amorphous systems with different sizes and disorder strengths. 
After tracing out the average topological invariant as a function of disorder for many system sizes, we collapse these curves by rescaling the disorder strength according to
\begin{equation}\label{eq_finsize_scaling}
\tilde{V} = V_c + (V - V_c) \left( \frac{L}{L_0} \right)^{1/\tilde{\nu}}.
\end{equation}
Here $L$ is the system size, $L_0$ is an intermediate size, $V_c$ is a critical disorder strength at which all curves should in principle intersect, and $\tilde{\nu}$ is the critical scaling exponent.
\Fig{fig_scaling} shows that both the Kitaev sum (top panels) and Bott index (bottom panels) collapse under finite-size rescaling of Eqn~\ref{eq_finsize_scaling} with the same values $V_c / \langle\Omega_p \rangle = 0.63 \pm 0.04$, and scaling exponent, $\tilde{\nu} \approx 2.6$, though small samples with fewer than $450$ gyroscopes deviate visibly from this scaling due to boundary effects in the Kitaev sum. 
This value is consistent with reported scaling exponents of $\tilde{\nu} = 2.58\pm 0.03$ found in previous studies of quantum tight binding models~\cite{xue_quantum_2013}.

Notably, we obtain reasonable fits for a surprisingly wide range of critical exponents $\tilde{\nu} = 2.2-3.0$, and a $\chi^2$ analysis constrains our value only within an uncertainty of $\Delta\tilde{\nu} \approx 1$.
Nonetheless, the scaling collapse suggests that the transition is, in fact, infinitely sharp in the limit of large system sizes, with finite size scaling consistent with the universality class of Cartan label A~\cite{barlas_topological_2018}. 
This result is consistent with the idea that amorphous Chern insulators share similar scaling behavior with tight binding models on lattices endowed with random pinning disorder~\cite{huo_current_1992,hung_disorder_2016}.
Our system is also different from the traditional crystalline case in another way in that gyroscopic metamaterials are symplectic rather than Hermitian.
It is interesting to note that neither difference appears to break the standard universal scaling~\cite{xue_quantum_2013}.
While two recent studies have found non-universal behavior in amorphous systems~\cite{ivaki_criticality_2020,sahlberg_topological_2020}, which motivates continued study of the scaling behavior along other directions of the phase boundary, we find no inconsistency with the scaling dictated by the symmetry class of the dynamical matrix.


\subsection{Topological Anderson Insulator phase diagram}

\begin{figure*}
\includegraphics[width=\textwidth]{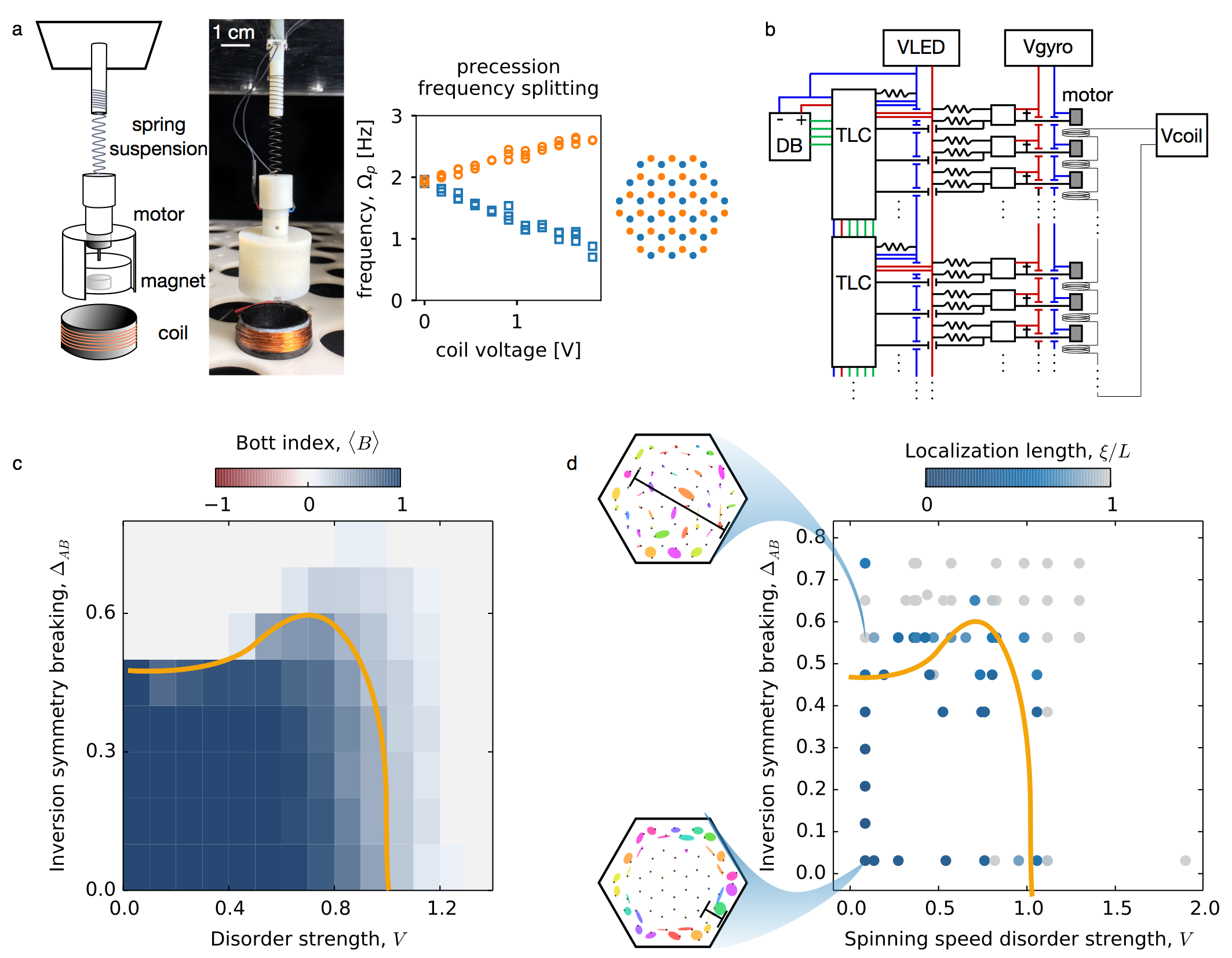}
\caption[]{
\textbf{
Gyroscopic networks exhibit a topological Anderson insulator (TAI) transition wherein a trivial insulator phase is driven into the topological phase by adding disorder.}
\textit{(a)} In our experiments, each gyroscope is constructed from an externally powered motor encased in a 3D printed housing that minimizes damping. 
A magnetic coil placed beneath each site allows variation of the on-site precession frequency $\Omega_p\rightarrow \Omega_p \pm \Delta_{AB}$, with each sublattice site (orange and blue) receiving an opposite bias from the magnetic torque.
\textit{(b)} Daisy-chained microcontrollers (TLCs) convert a serial output into a site-specific spinning speed via pulse width modulation, while a separate DC power supply drives inversion symmetry breaking by passing current through each coil.
\textit{(c)} A 2D topological phase diagram shows that either breaking inversion symmetry (increasing $\Delta$) or introducing strong disorder (increasing $V$) drives a transition from topologically nontrivial ($\langle B \rangle \ne 0$) to trivial ($\langle B \rangle =0$). 
Each colored region represents the average Bott index computed numerically for 200 random disorder configurations with nearest-neighbor interactions only. 
The pinning frequencies are chosen from a flat distribution of disorder of width $V = [ \max(\Omega_p) - \min({\Omega_p})] / \Omega_p^0$, and magnetic interaction disorder covaries with the pinning frequency, as would occur in the experiments. 
\textit{(d)} An experiment designed to realize the topological Anderson insulator transition in 54 gyroscopes measures the localization length of normal modes excited in a frequency range within the band gap of the clean system ($\Delta=0$, $V=0$). 
As either disorder strength or inversion symmetry breaking is increased, the localization length grows on average, reaching or exceeding the system size, $L$. 
Disorder is introduced by controlling the spinning frequency of each gyroscope individually in a spatially uncorrelated fashion, which varies both pinning and interaction disorder strength. 
The orange line is a guide to the eye taken from simulations. 
Though the experimental system exhibits many differences from the numerical idealization, including long range interactions and strong nutation at low spinning speeds, the qualitative features of the transition are evident.
 }
\label{fig_TAI}
\end{figure*}

We have seen that strong disorder destroys topological behavior.
By tuning an additional parameter, however, we find the remarkable possibility of transforming a trivial insulator into a Chern insulator simply by adding disorder to the system. 
Recently, this re-entrant phase diagram of so-called `topological Anderson insulation' was demonstrated for the first time in photonic~\cite{stutzer_photonic_2018} and cold atom systems~\cite{meier_observation_2018}.
As shown in~\Fig{fig_TAI}, gyroscopic metamaterials support this same transition.
Here, we simultaneously vary both random disorder strength and inversion symmetry breaking in a honeycomb lattice, both in experiment and in numerical calculations.
We introduce disorder by changing the gyroscope spinning speeds, which changes both pinning frequencies and bond strengths simultaneously due to the dependence of each on spinning speed.
Meanwhile, inversion symmetry is broken by a periodic array of staggered magnetic fields that splits the on-site precession frequencies of the two sublattice sites.

\Fig{fig_TAI}c shows the resulting topological index, here shown by the Bott index averaged over 200 realizations for each value of disorder. 
Increasing either inversion symmetry breaking or disorder strength drives the gyroscopic lattice into the trivial phase, but increasing both simultaneously allows for the topological phase to persist for significantly larger values of inversion symmetry breaking.
This enables a re-entrant phase transition by increasing disorder in a system with large inversion symmetry breaking.
Using the Kitaev sum in place of the Bott index gives the same result, aside from weakened convergence near the phase boundaries.
In~\Fig{fig_TAI}c, we choose $\Omega_k = 0.67 \Omega_p$ to match the relative values in our experiment, and also choose the disorder to affect both pinning and interaction strengths proportionally, as would occur in an experiment. 
The qualitative features of the resulting re-entrant phase transition, however, are indifferent to modest variations of these parameters, to the shape of the disorder distribution (flat versus Gaussian), and to whether or not interaction strength disorder is varied in tandem with pinning frequencies or instead of pinning frequencies.

\Fig{fig_TAI}d shows complementary measurements made in an experimental setup of 54 gyroscopes coupled with repulsive magnetic interactions.
Note that with magnetic interactions, the equations of motion are similar to the spring case considered thus far, as shown in Appendix~\ref{appendix:magnetic_interactions}. 
The edge of the sample is shaken at a slowly varying frequency that spans the band gap range of the clean ($V=0$) and inversion symmetric ($\Delta=0$) system.
Taking the Fourier transform of the tracked gyroscope displacements extracts the normal modes, from which we measure the localization length by fitting the displacement amplitudes to a decaying exponential as a function of distance from the edge.
The average localization length $\xi$ of modes in a narrow range of frequencies that lies in the gap of the clean system is reported as a fraction of the system width, $L$.
As the system approaches the boundary between topological and trivial phases, the localization length grows (gray dots in~\Fig{fig_TAI}d). 
In principle, the localization length should decrease again far from the phase boundary, though this subsequent decrease is not visible in our experiment except at low disorder.

In the experiments, the inversion symmetry is broken using a staggered array of magnetic coils, and the disorder strength is controlled by setting the spinning speeds of individual gyroscopes using a chain of microcontrollers (~\Fig{fig_TAI}b).
In our experimental setup, we set all experiments to have $\omega_0=175$ Hz spinning frequency, and we vary the width of the distribution using a pulse width modulation setup.
Briefly, we first set all gyroscopes to have identical spinning frequencies by iteratively adjusting the duty cycle (i.e., fraction of time that a gyroscope is receiving power) assigned to individual gyroscopes using a pulse width modulation controller. 
This step cancels out initial disorder from variations in the construction of our 3D-printed and hand-assembled gyroscopes (Stratasys Objet 350 printer).
We then generate a target spinning speed assignment for all gyroscopes so that the precession frequencies will approximately follow a flat distribution. 
We first attempt to achieve this assignment by perturbing the duty cycle values according to an average calibration relating duty cycle to spinning speed. 
We then iteratively measure the spinning speeds with a high-speed camera (Vision Research Phantom v12.1) and adjust the duty cycle values to approach the target spinning speed assignment.

There are several differences between the experimental realization and the idealized model.
For large disorder strengths, some gyroscopes are set to spin very slowly.
For these slowly spinning gyroscopes, nutation and nonlinear effects play a role in the dynamics, driving the system away from the fast-spinning-limit assumption. 
Additionally, the magnetic interactions are long range in the experiment. 
We choose to suppress both of these complications in the model shown in~\Fig{fig_TAI}c to underscore the simplicity of the phenomenon.

We use the localization length as an indication of the phase transition rather than measuring the Chern numbers of the phases themselves.
While this is an indirect measure of topology, the results capture the essential features of the expected phase diagram.

\section{Conclusion}
Periodic order is dispensable for generating topology in gyroscopic metamaterials. 
In light of this observation, by re-examining the fundamental dynamics of coupled gyroscopes, we have identified intuitive, real-space descriptions for the origins of topological gaps. 
The signatures of time reversal symmetry breaking -- an essential ingredient in endowing bands with non-trivial topology -- appear even at the three-gyroscope level through the interaction of counter-rotating polarizations. 
The three-way interaction between TRS breaking, network geometry, and coupled polarizations underlies the emergence of topological gaps, and these nontrivial gaps are topologically disconnected from bonding/anti-bonding energy splitting.
We present a method to predict the existence of a band gap without full diagonalization of the dynamical matrix that generalizes to amorphous structures. 
In the same spirit of simplifying the predictive machinery for topological phases irrespective of spatial order, we approximate the topological index using only local properties of a gyroscopic network, in contrast to other real-space methods in wide use. 
The ability to do so reflects the ability of even mesoscopic patches of gyroscopes to register as belonging to a topologically nontrivial phase, further supporting the notion that periodicity is not needed for nontrivial topology.

We find that disorder transitions in both amorphous and crystalline gyroscopic networks have similar behavior to periodic electronic and photonic Chern insulators. 
Amorphous networks display finite-size scaling that is consistent with the universal behavior from crystalline Chern insulators when subjected to pinning disorder on top of their existing structural disorder. 
Finally, we demonstrate a re-entrant topological phase diagram in a mechanical context for the first time.

\section{Acknowledgements}
Lisa M Nash contributed to this work via experimental design, gyroscope fabrication, conceptual insights, and useful discussions. 
We thank Emil Prodan for his extensive insight and Nikolas Claussen for a careful reading of the manuscript and helpful comments. 
We also thank Anton Souslov, Alex Edelman, Emmanuel Stamou, Tom Witten, and Daniel Hexner for useful discussions.
This work was supported in part by the University of Chicago Materials Research Science and Engineering Center, which is funded by National Science Foundation under award number DMR-1420709.
This research was supported in part by the National Science Foundation under Grant No.~NSF PHY-1748958.
NPM acknowledges support from the Helen Hay Whitney Foundation.
Additional support was provided by the Packard Foundation and by NSF EFRI NewLAW grant 1741685.


\appendix

\section{Gyroscopes with magnetic interactions}
 \customlabel{appendix:magnetic_interactions}{A}

If we replace the spring coupling with magnetic interactions as in ~\Fig{fig:intro_gyro_symplectic}c, the equations of motion are modified slightly in the experimental system, which uses magnetic interactions rather than springs.
In this case, the modified equation can similarly be cast in the form of~\Eq{eq:basicD} and appears as
\begin{equation}
	\label{eom}
    \begin{split}
	i \partial_t \psi_i = 
    \Omega_p \psi_i + \frac{1}{2} \sum_q 
    \bigg[ & \left(\Omega_{ii}^{+} \psi_i + \Omega_{ij}^{+} \psi_j\right)   \\
    & + e^{2 i \theta_{ij}}\left(\Omega_{ii}^{-} \psi^*_p 
+ \Omega_{ij}^{-} \psi^*_j\right) \bigg],
	\end{split}
\end{equation}
where the sum is over nearby gyroscopes, $\Omega^{\pm}_{ij} \equiv - \frac{\ell^2}{I\omega}\left( \partial F_{i\parallel} /\partial x_{j \parallel} \pm \partial F_{i\perp}/\partial x_{j\perp} \right)$
 is the characteristic interaction frequency between gyroscopes $i$ and $j$, 
$\Omega_p \equiv (mg + F^{\textrm{suspension}} + F^{\textrm{coil}}_z) \ell/I\omega$ is the precession frequency in the absence of other gyroscopes, 
and $\theta_{ij}$ is the angle of the bond connecting gyroscope $i$ to gyroscope $j$, taken with respect to a fixed global axis.
The interaction strengths, $\Omega^{\pm}_{ij}$, scale with the quantity $\Omega_k \equiv \ell^2 k_m/I\omega$, where $k_m$ is the effective spring constant for the magnetic interaction, and $\Omega^{\pm}_{ij}$ depend nonlinearly on the lattice spacing.

\section{Normal modes at the broken Dirac point for the gyroscopic honeycomb lattice}
 \customlabel{appendix:dirac_splitting}{B}

For the honeycomb lattice, we provide an argument for how symmetry determines the spatial structure of eigenmode displacements and their frequencies bounding the band gap.
The two states at a corner of the Brillouin zone have frequencies
\begin{equation}
\omega = 
\left\{
\begin{array}{c}
\pm \frac{1}{2}  (2 \Omega_p+3 \text{$\Omega $k}) 
\\ 
\pm \sqrt{\Omega_p (\Omega_p+3 \Omega _k)} 
\end{array}
\right\}.
\end{equation}
Using the basis $\mathbf{e} = (\psi^R_0, \psi^R_1, \psi^L_0, \psi^L_1)$, the positive frequencies have eigenvectors
\begin{equation}
\mathbf{e}^+ = \left\{
\begin{array}{c}
(1,0,0,0)
\\
\left(0,1,
\frac{- 3 \Omega_k}
{2 \Omega_p+3 \Omega_k +2 \sqrt{\Omega_p (\Omega_p+3 \Omega_k)}},0 \right) 
\end{array}
\right\}.
\end{equation}
For the pair of eigenvectors with larger absolute value of frequency, the excitations are solely $\psi^R$ for $\omega > 0$ or solely $\psi^L$ for $\omega < 0$. 
For the other normal mode pair with a smaller absolute value of frequency, the eigenvector is almost entirely in the opposite sublattice site, except for an excitation which is smaller in amplitude and opposite in chirality. 

We can see how translational and rotational symmetry enforces the modes' form at $K$ and $K'$ in detail.
First consider rotational symmetry of the lattice: 120$^\circ$ rotation preserve the momentum $K$ or $K'$ up to translation by a vector of the reciprocal lattice, 
so the modes at $K$ and $K'$ can be classified by their eigenvalue with respect to this operation.
This eigenvalue is a phase factor, $1$, $e^{i 2\pi / 3}$ or $e^{-i 2\pi /3}$.
Multiplication by a phase factor here corresponds to a displacement along the elliptical orbit of each gyroscope. 
~\Eq{eq:psiRpsiL} shows that multiplication by a phase, $\psi^R_p\rightarrow e^{i\theta}\psi^R_p$, $\psi^L_p\rightarrow e^{i\theta}\psi^L_p$, is equivalent to evolving in time by $\theta/\omega= \theta T/ 2\pi$, where $T$
is the period of the mode.
Hence rotational symmetry implies that
the displacement pattern of the gyroscopes returns to itself if the whole system is rotated by $120^\circ$ counterclockwise and then one waits by a time of $0$, $T/3$, or $-T/3$.
Consider specifically rotating $120^\circ$ around one of the gyroscopes. 
Any ellipticity in the gyroscope's path would prevent the possibility that the rotated configuration returns to itself after some time.
Thus, all nonzero gyroscope displacements must trace perfectly circular paths, either clockwise or anticlockwise.

Given that the orbits are circular, we can further constrain their pattern by considering the case of weak coupling compared to the pinning forces on each site ($\Omega_k \ll \Omega_p$).
In this limit, the components of motion that are clockwise have a large amplitude compared to the counterclockwise components.  
At the $K$ point, there are two modes, one in which the $A$ sites have a strong clockwise motion and one in which the $B$ sites have a strong clockwise motion.  
One could envision a mode in which all sites have a strong clockwise motion, but symmetry prohibits this as well, as we now show by considering one gyroscope moving with a small amplitude and the three gyroscopes around it moving with a large amplitude (\Fig{fig_diracmodes}b).

The relative phase of the three neighbors is determined by translational symmetry. 
Translating the excitation of an eigenmode by a lattice vector $\mathbf{R}$ merely introduces a phase shift of $T \mathbf{k} \cdot\mathbf{R}/2\pi$.
In~\Fig{fig_diracmodes}, $\mathbf{R}_1$ and $\mathbf{R}_2$ are the two primitive vectors of the lattice. 
Any gyroscope located $\mathbf{R}_1$ away has a position that lags behind by a time $T/3$, and any gyroscope located $\mathbf{R}_2$ away leads by a time of $T/3$.
Since these three gyroscopes are all moving clockwise, this means that in the first case the gyroscope will differ in phase by $2\pi/3$ counterclockwise from the reference gyroscope, and in the second it will be $2\pi/3$ clockwise.
If the gyroscope with a small clockwise amplitude is at an $A$ site (lower state in~\Fig{fig_diracmodes}a-b), then this implies that the displacements of the neighboring $B$-gyroscopes rotate counterclockwise, while if it is at a $B$ site, then the neighboring gyroscopes' displacements rotate clockwise as you go around the site clockwise (upper state in~\Fig{fig_diracmodes}a-b).  
As discussed in the text, this gives rise to a net force on the lower-frequency state and force balance on the immobile site of the higher-frequency state. 
A larger displacement at the counter-clockwise site therefore lowers the spring energy for the lower state, opening the gap as $\Omega_k / \Omega_p$ increases.

 \section{Proof of Upper bound of eigenvalues of a matrix}\customlabel{appendix:proof_gap}{C}
 
Here we provide an upper bound on the eigenvalues of a hopping Hamiltonian used in the main text; it relates their magnitude to  the sum of the hopping values from each site. To state this bound precisely, let $M$ be a Hermitian matrix. Let $A=\max_i \sum_j|M_{ij}|$.  Then if $\lambda$ is any eigenvalue of $M$, $|\lambda|\leq A$. Let $x$ be the eigenvector with
 eigenvalue $\lambda$.  Let $x$'s largest entry (in magnitude) be the $k^\mathrm{th}$ entry.  We have $\sum_j M_{kj}x_j=\lambda x_k$ by considering the $k^\mathrm{th}$ entry of $Mx=\lambda x$. So
 \begin{align}
 |\lambda||x_k|&\leq \sum_j |M_{kj}||x_j|\nonumber\\
 &\leq \sum_j |M_{kj}| |x_k|\nonumber\\
 &\leq A|x_k|.
 \end{align}
 The first line follows by the triangle inequality, the second by the assumption that $x_k$ is the largest entry in the vector and the third by the assumption that
 $A$ is the largest of the sums of the absolute values of entries in a row of $A$. Cancelling $|x_k|$ gives $\lambda\leq A$.
 
 \section{Exponential Decay of Green's Function}
 \customlabel{appendix:green_function}{D}
 
In section~\ref{Greens_func_yields_gap}, we argue that for a ordinary insulator, the Green's function expansion, which reads
\begin{equation}
    \frac{1}{\omega-D_{diag}}\sum_{n=0}^\infty \left(\delta   \frac{1}{\omega-D_{diag}}\right)^n,
\end{equation}
 decays exponentially with distance. 
 Here, we provide more detail on this point.
 The largest contribution to the geometric series for the Green's function is from the shortest path connecting a given pair of sites.  However, the main contribution
 to the Green's function could in principle be from longer paths, because although longer paths give a smaller contribution to the sum, there are many more paths that meander through many sites from site $i$ to $j$ than direct paths. 
 To give a rough (but rigorous) upper bound that takes this into account, note that there cannot be more than $z^n$ paths
 of length $n$ starting at site $i$ if each site has $z$ connections.
 In particular, there are not more than $z^n$ paths connecting site $i$ to site $j$ in $n$ steps.
Therefore, the largest the Green's function can be is $\sum_{n=m}^\infty(z \alpha /\beta)^n$, where $\alpha=\max_{pq} |{\delta_{pq}}|$ is the maximum off-diagonal entry of the dynamical matrix, and $\beta=\min_p|\omega-D_{pp}|$ is the minimum on-diagonal component of $\omega-D$. 
This sum converges to a constant times $(z\alpha /\beta)^n$ if the off-diagonal elements are small enough.  
This shows that the Green's function decays exponentially.
A similar argument can be used to show that the Green's function and projection operator are also localized for the topological insulator case based on the expansions in Equations~\ref{eq:green_series} and~\ref{eq:projector_series_expansion}.
 
 \section{Local approximation without nearly diagonal $D^2$}\customlabel{appendix:generalize_chern}{E}

When a system has a dynamical matrix such that $D^2$ is nearly diagonal, it is possible to calculate its Green's function (within a gap) and the Chern number in a local way (see Eqs.~\ref{eq:green_series}, \ref{eq:projector_series_expansion}).
This expansion works also when $D^2$ is not close to a scalar, as long as
the system has a gap (or probably even just a mobility gap).
If one wants to find an eigenvalue close to a certain value, one can shift the diagonal entries of the matrix uniformly so that this value is at zero, square the matrix, and then use the Lanczos algorithm to find the minimal eigenvalue of the squared matrix~\cite{lanczos_iteration_1950,ojalvo_vibration_1970}.  
In a similar way, we can find the Green's function near a given frequency. 
To calculate
$(\omega\mathds{1}-D)^{-1}$, square $\omega\mathds{1}-D$ and then subtract a scalar, $\omega_0^2\mathds{1}$, to make what remains as small as possible: $(\omega\mathds{1}-D)^2=\omega_0^2\mathds{1}+\epsilon$. 
For a dynamical matrix whose square is nearly a scalar, $\epsilon$ is extremely small.  
However, it is possible that a dynamical matrix does not nearly square to a scalar, and here we generalize our earlier result to this case. 

Physically, if $
\epsilon=\left(\frac{(\omega\mathds{1}-D)^2-\omega_0^2\mathds{1}}{\omega_0^2}\right)$ has large entries, then interference effects fail to keep an excitation localized within only two steps. 
Yet interference may arise between sufficiently long paths; correspondingly, the entries of higher powers of $\epsilon$ start to decay exponentially. 
One can then expand the Green's function, 
 $(\omega-D)^{-1}=(\omega\mathds{1}-D)(\omega_0^2\mathds{1}+\epsilon)^{-1}$ by using a geometric series to calculate the reciprocal.
This \emph{always} occurs for the right choice of $\omega_0$, as long as $\omega$ is in a gap. 

The Chern number of the states above a frequency $\omega$ can be calculated in the same way. The Kitaev sum depends on the projection, which we express as
 $P=\frac{1}{2}\mathds{1}+\frac12\frac{\omega\mathds{1}-D}{\sqrt{(\omega\mathds{1}-D)^2}}$.  
 The square root can be expanded just as the Green's function was (the radii of convergence of the binomial expansion of
$\sqrt{\omega_0^2+x}$ and of $(\omega_0^2+x)^{-1}$ are identical), giving
\begin{equation} 
P=\frac{1}{2}\mathds{1}+\frac{\omega-D}{2\omega_0}\sum_{n=0}^\infty\left(-\frac{\epsilon}{4\omega_0^2}\right)^n \binom{2n}{n}.
\end{equation}

Let us argue that there is always a value of $\omega_0$ such that this converges.  
The eigenfrequencies of $D$ are contained in two intervals,
$[M_1,\omega-\Delta_1]$, and $[\omega+\Delta_2,M_2]$, where $\Delta_1,\Delta_2$ are the distances to the edges of the bands, and $M_1, M_2$ are the minimal and maximal frequencies.  Thus the eigenvalues of $(\omega-D)^2$ are all contained between $\kappa=\mathrm{min} (\Delta_1^2,\Delta_2^2)$ and $K=\mathrm{max}((M_2-\omega)^2,(\omega-M_1)^2)$. Let $\omega_0^2$ be the center of the spectrum, $\frac{1}{2}(\kappa+K)$, and define $\epsilon$ as above.
Then $\epsilon$'s eigenvalues are between $\pm\frac12(K-\kappa)$, which is less than $\omega_0^2=\frac12 (K+\kappa)$ because $\kappa>0$, and thus the expansion converges. 

This argument has thus far relied on computing the range of eigenvalues.
However, our goal is to calculate the projection and the Chern number without diagonalizing the dynamical matrix.  
To connect to our aim, we can note that if the terms of the series decay exponentially in the basis of eigenvectors, then they also decay exponentially in real space.  
One can find a useful value for $\omega_0$ by calculating the terms of the series in real space for different $\omega_0$'s until one finds that it converges.

The argument above shows that the series must converge so long as there is a gap.
In fact, large values of $\omega_0$ always lead to a convergent series if a gap exists: in particular, values of $\omega_0$ such that $\omega_0^2>\frac12 K$ guarantee convergence if a gap exists. 
As a consequence, checking for convergence with $\omega_0^2>\frac12 K$, where $K$ is any upper bound on the spectrum, immediately signals whether there is a gap (in which case the series converges) or there is no gap (in which case the series does not converge). 

The limitation of this approach is that the convergence can be slow, requiring a numerical approach.
Slow convergence could result if the bands are wide compared to the gap or if there are many other bands besides the two bands bounding the gap.  
This is the case, for example, for gyroscopic lattices, in which there is a particle-hole symmetry that results in an asymmetric distribution of bands around any positive-frequency gap. 
In such situations, the convergence improves if one applies a polynomial to $D$ to fold the spectrum so that the bands on each side of $\omega$ bunch together in frequency.

\bibliography{refs}
\end{document}